\documentclass{aastex61}
\usepackage{graphicx}
\usepackage{amsmath}
\usepackage[flushleft]{threeparttable}
\usepackage{color}
\usepackage{gensymb}
\usepackage{booktabs}

\usepackage{siunitx}
\usepackage{rotating}
\usepackage[utf8]{inputenc}
\newcommand\aastex{AAS\TeX}

\newcommand{\beq}{\begin{equation}}
\newcommand{\eeq}{\end{equation}}

\shorttitle{\aastex\ Stellar parameters of final release of MaStar}
\shortauthors{Chen et al.}

\begin{document}


\title{SDSS-IV MaStar: Determination of Stellar Parameters Using Bayesian Averaging}

\author{Yan-Ping Chen}
\affil{Center for Astrophysics and Space Science (CASS), New York University Abu Dhabi, P.O. Box 129188,  Abu Dhabi, United Arab Emirates\email{chenyp.astro@gmail.com}}

\author{Renbin Yan}
\affiliation{Department of Physics,The Chinese University of Hong Kong, Shatin, N.T., Hong Kong S.A.R., China\email{ rbyan@cuhk.edu.hk}}

\author{Szabolcs~M{\'e}sz{\'a}ros} 
\affiliation{ELTE E\"otv\"os Lor\'and University, Gothard Astrophysical Observatory, 9700 Szombathely, Szent Imre H. st. 112, Hungary}
\affiliation{MTA-ELTE Lend{\"u}let ``Momentum" Milky Way Research Group, Hungary}
\affiliation{MTA-ELTE Exoplanet Research Group, Hungary}

\author{Claudia Maraston}

\affiliation{Institute of Cosmology \& Gravitation, University of Portsmouth, Dennis Sciama Building, Portsmouth, PO1 3FX, UK}

\author{Daniel Thomas}
\affiliation{Institute of Cosmology \& Gravitation, University of Portsmouth, Dennis Sciama Building, Portsmouth, PO1 3FX, UK}
\affiliation{School of Mathematics and Physics, University of Portsmouth, Lion Gate Building, Portsmouth, PO1 3HF, UK}

\author{Daniel Lazarz}
\affiliation{Department of Physics and Astronomy, University of Kentucky, 505 Rose St., Lexington, KY 40506-0057, USA}

\author{Guy S. Stringfellow}
\affiliation{Center for Astrophysics and Space Astronomy, University of Colorado, 389 UCB, Boulder, CO 80309-0389, USA} 

\author{Lewis Hill}
\affiliation{Institute of Cosmology \& Gravitation, University of Portsmouth, Dennis Sciama Building, Portsmouth, PO1 3FX, UK}

\author{Julie Imig}
\affiliation{Department of Astronomy, New Mexico State University, Box 30001, MSC 4500, Las Cruces NM 88003, USA}

\author{Joseph D. Gelfand}
\affil{Center for Astrophysics and Space Science (CASS), New York University Abu Dhabi, P.O. Box 129188,  Abu Dhabi, United Arab Emirates}

\author{Jon A. Holtzman}
\affiliation{Department of Astronomy, New Mexico State University, Box 30001, MSC 4500, Las Cruces NM 88003, USA}

\author{Matthew {Bershady}}
\affiliation{University of Wisconsin - Madison, Department of Astronomy, 475 N. Charter Street, Madison, WI 53706-1582, USA}
\affiliation{South African Astronomical Observatory, PO Box 9, Observatory 7935, Cape Town, South Africa}
\affiliation{Department of Astronomy, University of Cape Town, Private Bag X3, Rondebosch 7701, South Africa}

\author{Dmitry Bizyaev}
\affiliation{Apache Point Observatory and New Mexico State University, P.O. Box 59, Sunspot, NM 88349, USA}
\affiliation{Sternberg Astronomical Institute, Moscow State University, Universitetskij pr. 13, Moscow, Russia}

\author{Niv Drory}
\affiliation{McDonald Observatory, The University of Texas at Austin, 1 University Station, Austin, TX 78712, USA}

\author{Keivan G. Stassun}
\affiliation{Department of Physics and Astronomy, Vanderbilt University, VU Station 1807, Nashville, TN 37235, USA}







\begin{abstract}
We  present the stellar parameters for 
 59,266 high quality spectra of 24,130 unique stars
{\footnote{The `good visits' refers to a subset of visits with high-quality spectra. This excludes those spectra affected by extinction issues in their flux calibration, those that are marked as bad in visual inspection, or those with median S/N per pixel less than 15. This includes objects that contain emission lines, objects that have unreliable radial velocity measurements, and objects that may be affected by scattered light. The associated spectra of `Good Visits' are referred as `Good Spectra' \citep{Yan19}.}}
from the MaNGA Stellar library (MaStar) in the SDSS DR17 data release \citep{Abdurro'uf22}. 
The median signal-to-noise (S/N) ratio per pixel of the spectra is 96. We derive four stellar parameters, effective temperature ($\rm T_{eff}$), surface gravity ($\log g$), metallicity ($\rm[M/H]$) and alpha-enhancement ratio ($\rm [\alpha/M]$) 
by comparing the data with 
BOSZ \citep[ATLAS-9 based;][]{BOSZ} and MARCS \citep{Gustafsson08} theoretical atmospheric models.
We adopt a Bayesian method and use color and absolute magnitude derived from Gaia to select a subset of theoretical models for each star. We then perform full-spectrum fitting to estimate the likelihood of each model in the subset and then compute their likelihood-weighted mean parameters as the final parameters. 
We set stellar parameter quality flags to facilitate the use of 
the derived stellar parameters. 
The MaStar stellar parameters derived herein 
span an effective temperature range of
 $\rm 2, 600 \leq T_{eff} \leq 29, 861 K$, 
a surface gravity range of $0 \leq \log g \leq 5.5$, a metallicity range of $-4.9 \leq \rm[M/H] \leq 1.0$,  and an  $\alpha$-abundance range of $\rm  -0.98 \leq [\alpha/M] \leq 1.0$.
We compare these parameters with those from APOGEE and Gaia for stars in common, finding general consistency within the uncertainties. However, some artifacts and systematic differences are present, and we discuss their potential causes.
These new stellar parameters are available through the MaStar SDSS-IV DR17 link to our value-added catalog{\footnote{\url{https://www.sdss4.org/dr17/mastar/mastar-stellar-parameters/}}}.




\end{abstract}



\section{Introduction} \label{sec:intro}

Stellar libraries serve essential roles in a wide range of astrophysics applications. In particular, they are 
fundamental inputs for models of stellar populations. 
A stellar library is a collection of spectra from a variety of stars that cover a certain parameter space of atmospheric properties. 
There are two major categories of stellar libraries: theoretical stellar libraries that are built from atmospheric models assuming certain line lists and thermodynamic equilibrium status, and empirical stellar libraries that collect the spectra from observations.  
Examples of theoretical stellar libraries include, but are not limited to,
\citet{kurucz79, Zwitter04, Martins05, Munari05, Coelho05, Coelho07, Gustafsson08, Leitherer10, deLaverny12, BOSZ}.
Examples of empirical libraries include MILES \citep{milesref}, Pickles \citep{Pickles85, Pickles98}, 
\citet{Diaz89}, \citet{Silva92}, Lick/IDS \citep{Worthey94}, \citet{Lancon2000}, STELIB \citet{stelibref}, ELODIE \citet{elodie}, INDO-US \citep{Valdes04}, CaT \citep{Cenarro01}, MILES \citep{milesref,Falcon-barroso11}, HST NGSL \citet{ngsl}, X-Shooter Stellar Library \citep[XSL,][]{XSL,dr3xsl}, the NASA Infrared Telescope Facility (IRTF) Library \citep{Rayner09}, the Extended IRTF library \citep{Villaume17}, and the MUSE library \citep{Ivanov19}. 

Our team, motivated by the need to model the 
spectra of galaxies observed as part of the Mapping Nearby Galaxies at Apache Point Observatory \citep[MaNGA,][]{Bundy15, Yan16} survey, carried out the MaNGA stellar library  project \citep[MaStar;][]{Yan19} aiming to build a large, comprehensive stellar library that includes stars covering 
as wide a stellar parameter space as possible.
The stellar parameters deemed fundamental as a set
are effective temperature ($\rm T_{eff}$), surface gravity ($\log g$), metallicity ($\rm [M/H]$) and $\alpha$-abundance ($\rm [\alpha/M]$). With the parameters, a stellar library together with stellar tracks and isochrones can be used to build stellar population models \citep[e.g.,][]{BC03, Maraston05}. Our team has explored in detail various independent methods to derive stellar parameters, which include presented in \citet{Chenparam20, Hillmainparam,  Hillalphaparam, Imigparam, Lazarz22}, and this work.


Our results are inferred from the comparison between the data and the template models. Due to our extensive coverage of stellar parameter, theoretical atmospheric models are used. To minimize possible systematic errors introduced by interpolation between model grid points, 
a large, fine grid with small increments in stellar
parameters was constructed
at the cost of significant computation time, thereby bypassing the need for interpolation. 
In Section~\ref{secdata} we briefly review the MaStar data. In Section~\ref{sectemp}, we present the template features used in this work. In Section~\ref{secmeth}, we describe the method used to determine
the stellar parameters. The validation method is discussed in Section~\ref{secval}. Our results are presented in Section~\ref{secresul}.
Stellar parameter quality control is discussed in Section~\ref{secqc}, and consistency with parameters in the literature 
is presented in Section~\ref{secconsis}.
We conclude in Section~\ref{secconclusion}.\\

\section{data}\label{secdata}

The final release of data \citep{Yan24} follows the same strategy of observations and methodology as the first data release of MaStar \citep{Yan19}. We briefly summarize them here.
The majority of MaStar spectra are collected using the MaNGA fiber bundles in tandem with APOGEE-2N observations during APOGEE-led bright-time observations. 
Observations were carried out on the 2.5-meter telescope \citep{Gunn06} at Apache Point Observatory. We use the same spectrographs \citep{Smee13} and fiber feed system \citep{Drory15} as the MaNGA survey. 
The reduction of MaStar spectra is performed using the MaNGA Data Reduction Pipeline \citep[DRP;][]{Law16} with some modifications made to extract stellar spectra. The modification is described in the DR17 paper \citep{Abdurro'uf22}.
The MaStar spectra span a wavelength from 3622 \AA\  to 10354 \AA\ with a mean resolving power of $R \sim 1800$, although the resolution varies some with wavelength,
i.e., line-spread-function (LSF) \citep{Law21}.

\begin{figure}
\centering
\includegraphics[scale=0.4,angle=0]{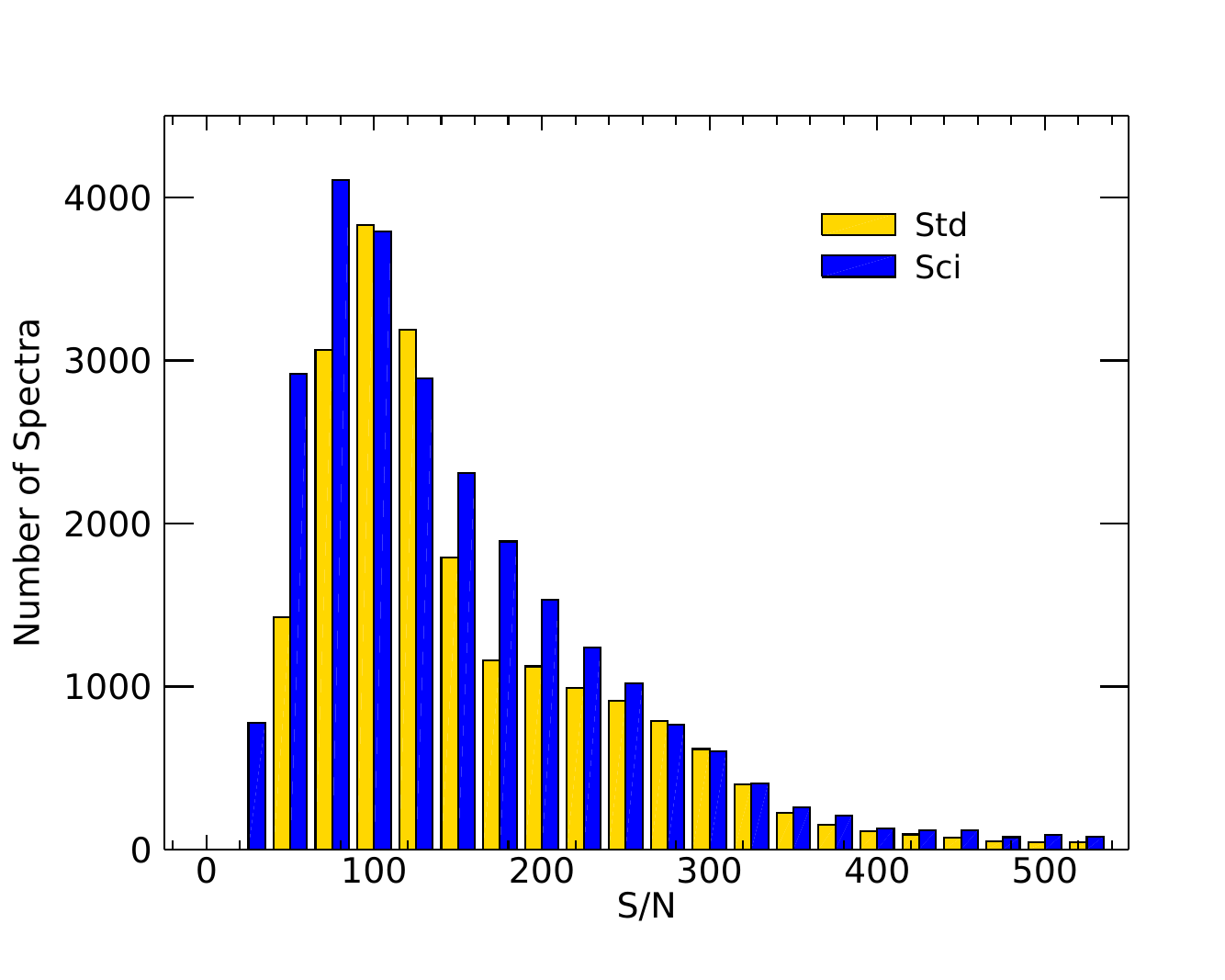}
\caption{Median signal-to-noise (S/N) per pixel distribution of the MaStar spectra. Flux standard stars (yellow) peak at S/N$\sim$100, science stars (blue) peak at S/N$\sim$ 80 with a wider distribution.}
\label{fig:sndistri}
\end{figure}

The subset of spectra used in this work is drawn from the final release of MaStar \citep{Yan19, Yan24} in SDSS DR17 \citep{Abdurro'uf22}. 
This release contains  11817 unique science stars and 12345 unique spectrophotometric standard stars, with an overlap of 32 stars targeted for both reasons. Each star was observed on one or more than one nights which we term as ``visits". A ``good visit" refers to a night of observation that produced a good quality spectrum for a given star. The quality selection criteria for `good quality spectra' and `good visits' are described in detail by \citet{Yan19} and updated by \citet{Abdurro'uf22}.
MaStar contains 59,266 good quality spectra for 24,130 unique stars. 
The distribution of median signal-to-noise ratio per pixel of the spectra is shown in Fig.~\ref{fig:sndistri}. The median of the distribution is 109.
The flux calibration of our data is accurate to $\sim4\%$ \citep{Yan19}. For further details on data reduction, we refer readers to \cite{Yan19,Yan24} and \citet{Abdurro'uf22}.



\section{Stellar Atmosphere Templates}\label{sectemp}
\subsection{The New BOSZ Grid}\label{secbosz}

\begin{figure}[!ht]
\epsscale{0.75}
\plotone{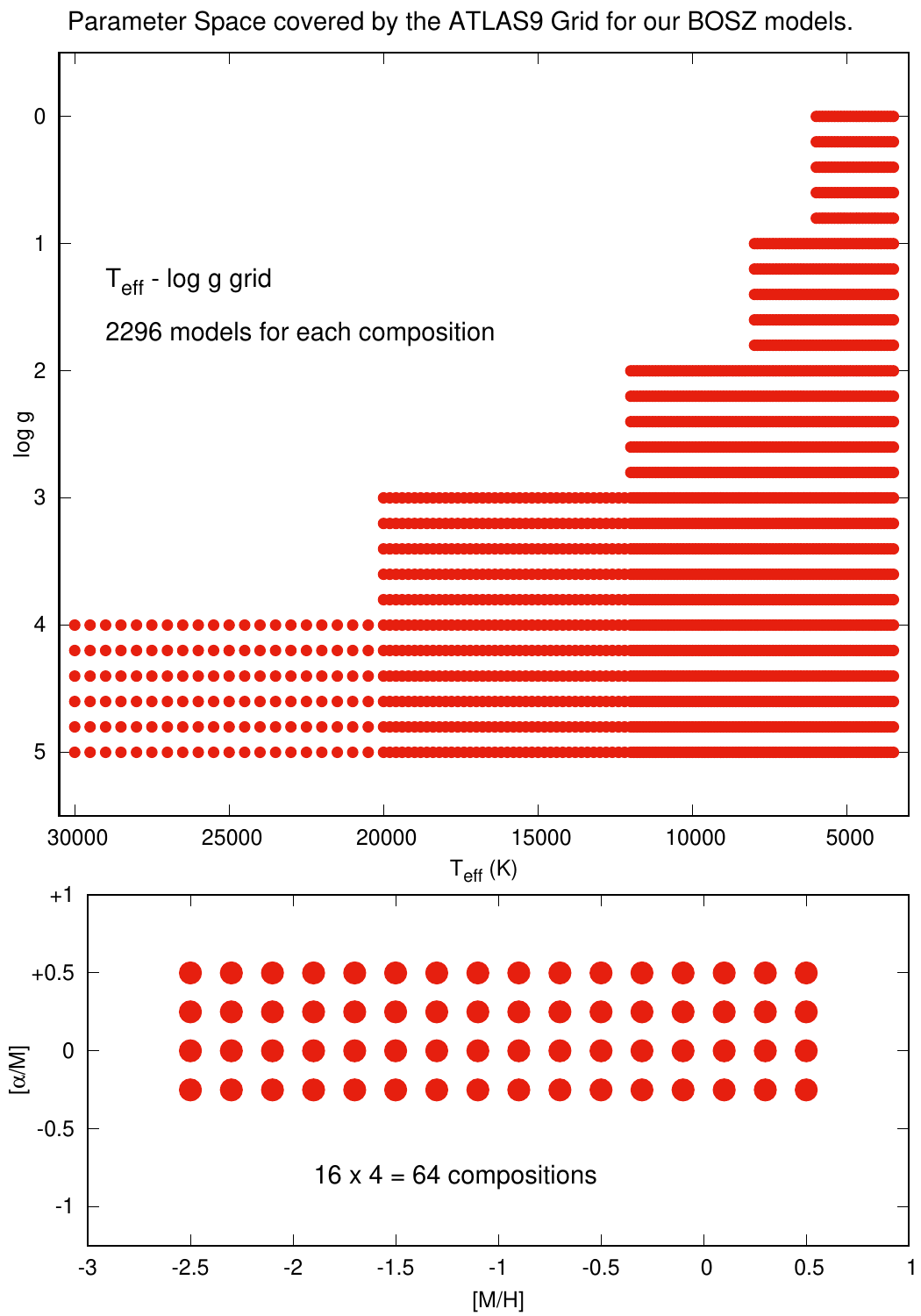}
\caption{Top panel: The T$_{\rm eff} - \log$~g space of model atmospheres used for synthesis for BOSZ models. 
Bottom panel: Grid resolution for  [$\alpha$/M]  as a function of metallicity for BOSZ models. }
\label{fig:synspace}
\end{figure}

\begin{figure}[!ht]
\epsscale{0.51}
\plotone{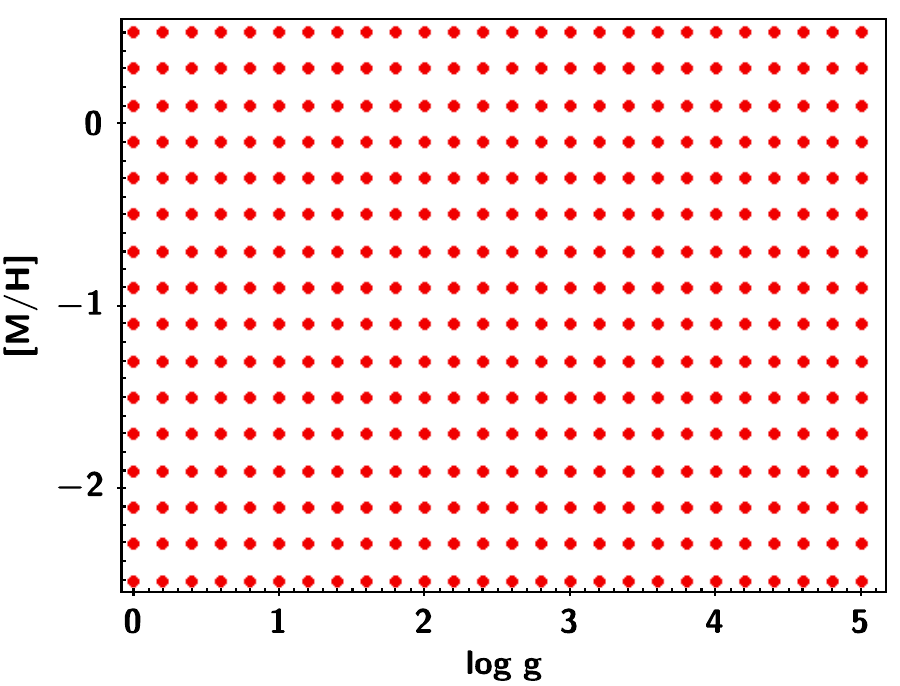}
\plotone{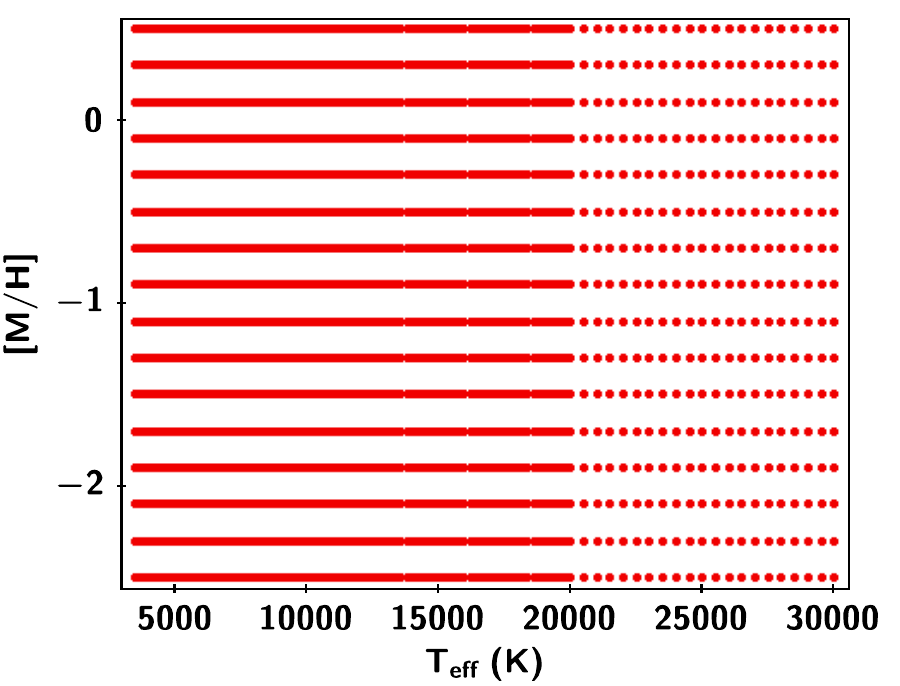}
\caption{ Left panel:  The T$\rm_{eff} - [M/H]$ space of model atmospheres used for synthesis for BOSZ models. Right panel:  The $\rm \log~g - [M/H]$ space of model atmospheres used for synthesis for BOSZ models. }
\label{fig:synmetal}
\end{figure}

\begin{deluxetable}{lrrrlrrr}
\tablecaption{Atmospheric Parameter grid for ATLAS9 Theoretical Spectra \label{tabboszgrids}}
\tablewidth{0pt}
\tablehead{
Parameter & \colhead{Min} & \colhead{Max}    & \colhead{Step}  &  
Parameter & \colhead{Min} & \colhead{Max}    & \colhead{Step}
}
\startdata
$[$M/H$]$ & $-$2.5 & 0.5 & 0.2 & & & &\\
$[$C/M$]$ & 0.0 & 0.0 & 0.0 & & & & \\
$[\alpha$/M$]$ & $-$0.25 & 0.5 & 0.25 & & & & \\
T$_{\rm eff}$  & 3500 & 6000 & 50 & $\log~g$ & 0 & 5 & 0.2 \\
T$_{\rm eff}$  & 6100 & 8000 & 50 & $\log~g$ & 1 & 5 & 0.2 \\
T$_{\rm eff}$  & 8100 & 10000 & 50 & $\log~g$ & 2 & 5 & 0.2 \\
T$_{\rm eff}$  & 10100 & 12000 & 100 & $\log~g$ & 2 & 5 & 0.2 \\
T$_{\rm eff}$  & 12200 & 20000 & 200 & $\log~g$ & 3 & 5 & 0.2 \\
T$_{\rm eff}$  & 20500 & 30000 & 500 & $\log~g$ & 4 & 5 & 0.2 \\
\enddata
\end{deluxetable}

In the age of large spectroscopic surveys, e.g., APOGEE-2 \citep{Majewski17}, Gaia-ESO \citep{Gilmore22} and \citep{Randich22}, GALAH \citep{deSilva15}, LAMOST \citep{cui12, deng12, zhao12},  WAVES \citep{Driver19}, 4MOST \citep{dejong19}, MOONS \citep{Cirasuolo14}, WST \citep{Mainieri24},
the need to have
a publicly available large spectral database calculated using the same software and consistent 
abundances became clear. Before these surveys, different spectral grids were (and still are) available in the literature; for a 
complete list, see the introduction by \citet{coelho14}. These older ATLAS9 grids were all synthesized by using 
the solar reference abundance table from either \citet{grevesse98}, 
or \citet{anders89}, and were limited to only a handful of compositions, insufficient to
cover the large parameter range these 
surveys require. In the early 2000s, significant changes and improvements were made to the solar composition 
table \citep{asplund05, grevesse07, asplund09}, which made it necessary to calculate grids of stellar spectra using 
an updated solar reference table. 

This effort was led by \citet{meszaros12} who carried out model atmosphere calculations using ATLAS9 \citep{kurucz79} 
for a much larger number of compositions than previously published. This grid covered metallicities 
([M/H]) from [M/H]=$-5$ to [M/H]=1.0 with varying carbon ([C/M]) and $\alpha$ ([$\alpha$/M]) abundances 
from $-$1.5 to +1.0 relative to metallicity using 2 $km\ s^{-1}$ micro-turbulent velocity with steps of 0.25~dex for each element. 
The $\alpha$-elements being varied in these calculations were: O, Ne, Mg, Si, S, Ca, and Ti. This grid was 
calculated by using solar abundances from \citet{asplund05}, which was a significant update from the previous grids. 
While newer solar abundance references are now available \citep{grevesse07, asplund09}, the changes are small and do not affect 
the structure of model atmospheres. This model atmosphere database can be downloaded from the 
ATLAS-APOGEE website\footnote{\url{http://www.iac.es/proyecto/ATLAS-APOGEE/}}.

In order to match the dense parameter distribution of MaStar, we chose to update parts of the model atmosphere database of  \citet{meszaros12} by re-calculating the models, namely, decreasing the step size to 0.2 dex in metallicity between $\rm [M/H]=-2.5$ and 0.5 when $\rm [C/M]=0$, and to
100~K in effective temperature when 
T$_{\rm eff} <$12000~K. The model atmosphere with T$_{\rm eff} \geq $12000~K are taken from \citet{meszaros12}. The selected atmospheric parameters are listed in Table~\ref{tabboszgrids}, and shown in Figs.~\ref{fig:synspace} and ~\ref{fig:synmetal}.
We kept [C/M]=0 and allowed $[\alpha/M]$ to vary between 
$-$0.25 and 0.5 in 0.25~dex steps. Hence, the new finer grid is intended to cover the main parameter range of MaStar where the vast majority of stars lie. 
In order to be consistent with the older grid, we used the same line lists and 
followed the same steps by using the same scripts developed by \citet{meszaros12} when computing the new model 
atmospheres. Convection was turned on with the mixing-length parameter set to $l/Hp$ = 1.25, but convective
overshooting was turned off. All the parallel model atmospheres have the same 72 layers from log $\tau_{\rm Ross}$ =  - 6.875 to 2, where 
the step size is 0.125. We refer the reader to \citet{meszaros12} for the exact details of the calculation procedure and 
the set convergence criteria. The new final grid consists of 64 compositions, 2296 models in each, totaling 146,944 model 
atmospheres. 

This new atmosphere dataset was the underlying model atmosphere grid for a new spectral 
synthesis to create a spectral library used by MaNGA. 
The spectral synthesis followed the same steps as 
\citet{bohlin17} in order to be consistent with the BOSZ spectral grid\footnote{\url{https://archive.stsci.edu/prepds/bosz/}}. 
The high-resolution spectra were calculated with SYNTHE \citep{kurucz81} using the Linux-ported version \citep{sbordone04}.  
The spectral grid was computed for the new parameter grid discussed above.
The spectra span a 
wavelength range between 300nm and 11 microns using vacuum wavelengths and were first synthesized with a resolution 
of 300,000 without convective overshooting and with a mixing length parameter of 1.25. 
Unlike the original BOSZ grid \citet{bohlin17} we chose to compute these for only one resolution of 10000, 
which we convolved the high-resolution spectra to to save computation time. 
The computed fluxes were sampled evenly in the logarithmic wavelength space, and we sampled each spectra the same 
way for each resolution. 
The spectra contain no rotational broadening, while the micro-turbulent velocity was used in the model atmospheres as 2 $km\ s^{-1}$. 

The line lists used are also the same as for the BOSZ grid \citep{bohlin17}. The atomic line list adopted was 
compiled by Robert Kurucz\footnote{\url{http://kurucz.harvard.edu/linelists.html}}. 
This line list was used without any modification by us, and was complemented with the following molecular line 
lists: H$_{2}$O, CH, MgH, NH, OH, SiH, H$_{2}$, C$_{2}$, CN, CO, SiO, and 
TiO\footnote{\url{http://kurucz.harvard.edu/molecules.html}}. The H$_{2}$O  \citep{partridge97} and TiO \citep{schwenke98}
line lists were formatted by Robert Kurucz so that they are compatible with ATLAS. 
Water was included for stars cooler than 
5500~K, while TiO only entered the calculations below 4500K to reduce computation time for temperatures where these 
molecules do not appear in the spectra. 

\subsection{MARCS model at cooler regions}

As the MaStar sample extends to cooler ($\rm T_{eff} \leq 3500K$) temperatures \citep{Yan19}, we use the MARCS\footnote{\url{https://marcs.astro.uu.se/}} \citep{Gustafsson08} atmospheric models in this stellar parameter region, as done by \citet{Hillmainparam,Hillalphaparam}. MARCS is a grid of one-dimensional, hydrostatic, plane-parallel, and spherical LTE model atmospheres. We use the MARCS models that were computed specifically for the APOGEE project \citep{Holtzman18, Jonsson20}. 

We choose to use the MARCS models at a resolution R=20,000 with $\rm T_{eff} \leq 5500\ K$ for the cool end of parameter space (especially below 3500K) for the MaStar parameter determination. 
The MARCS stellar atmospheric models used in this work are in the range of $\rm 2500K \leq  T_{eff} \leq 5500K$ in steps of 100K from $\rm 2500K  \leq   Teff  \leq  4000K$, and 250K between 4000K--5500K. The surface gravity spans $\rm -0.5  \leq   \log g  \leq  5.5$   in steps of 0.5 dex. The metallicity coverage spans $\rm -5.0 \leq [M/H] \leq +1.0$, in steps of 0.25 dex from $\rm -0.5 \leq [M/H] \leq +1.0$, 0.5 dex between $\rm -3.0 \leq [M/H] \leq  -0.5$, and 1.0 dex between $\rm -5.0  \leq [M/H] \leq  -3.0$. The metallicity grid is the same as in \citet{Jonsson20}.
As noted by \citet{meszaros12}, MARCS models use different molecular lines, such as $\rm H_{2}O$ and $\rm TiO$ compared to those used in the ATLAS based BOSZ models at cool temperatures. We refer to \citet{Gustafsson08} and \citet{Jonsson20} for details on the atomic and molecular line lists used in the MARCS models. 
The model atmospheric parameters of the MARCS grid used in this work are listed in Table~\ref{tabmarcsgrids}. 
The distribution of the MARCS model grid in atmospheric parameters space is presented in \citet{Jonsson20} in their Fig. 2.

\begin{deluxetable}{lrrclrrr}
\tablecaption{Atmospheric Parameters of MARCS Spectra  \label{tabmarcsgrids}}
\tablewidth{0pt}
\tablehead{
Parameter & \colhead{Min} & \colhead{Max}    & \colhead{Step}  &  
Parameter & \colhead{Min} & \colhead{Max}    & \colhead{Step}
}
\startdata
$[$M/H$]$ & $-$5.0 & 1.0 & variable  & & & &\\
$[$C/M$]$ & 0.0 & 0.0 & 0.0 & & & & \\
$[\alpha$/M$]$ & $-$1.0 & 1.0 & variable & & & & \\
T$_{\rm eff}$  & 2500 & 4000 & 100 & $\log~g$ & -0.5 & 5.5 & 0.5 \\
T$_{\rm eff}$  & 4000 & 5500 & 250 & $\log~g$ & 0.0 & 5.5 & 0.5 \\
\enddata
\end{deluxetable}

\section{Method}\label{secmeth}

The stellar parameters for the MaStar stellar library are determined by comparing the model spectra with observed spectra. For each
observed spectrum, full-spectrum-fitting is performed
using the ULySS \citep{Koleva08} code\footnote{\url{http://ulyss.univ-lyon1.fr/}} code, 
which generates a $\chi ^2$ value describing how well the model spectrum represents the observed spectrum as in \citet{Chenparam20}.
The spectrum-fitting using both the BOSZ and MARCS models is carried out separately 	
without intra-grid interpolation. 
Due to processing time constraints, we do not compare each MaStar spectra with all model spectra;  instead, we use priors to pre-select a subset of models for comparison. We describe these priors below in Section \ref{prior}.
The templates are selected near the priors within a box with a fixed size in the parameter space.

\subsection{Priors of the stellar parameters}\label{prior}

\begin{figure}
\centering
\includegraphics[scale=0.60,angle=0]{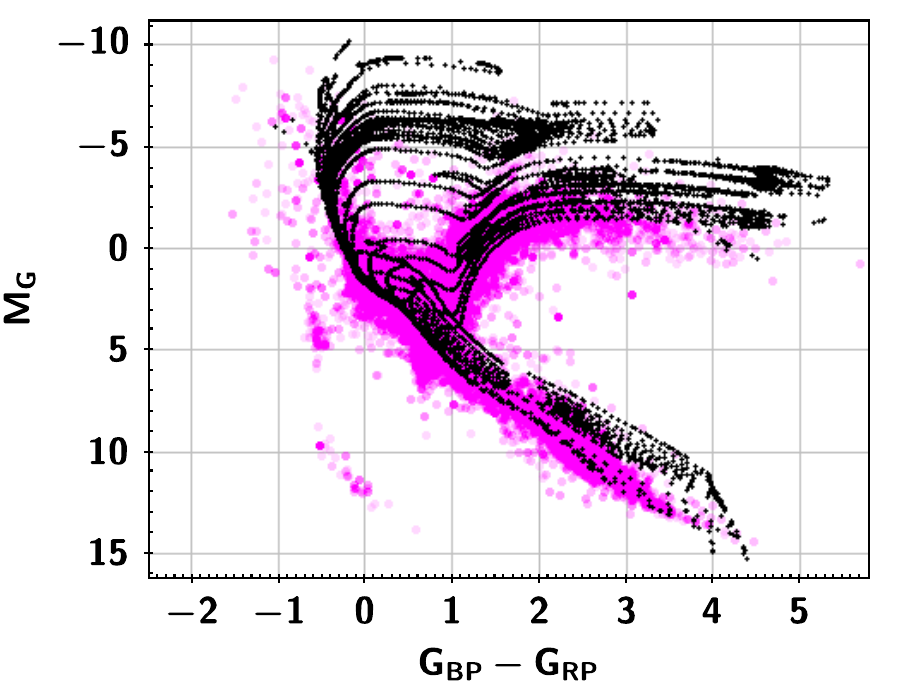}
\caption{Gaia color-magnitude diagram for MaStar library stars (pink dots) identified as ``good". Colors are corrected with the same algorithm presented in \citet{Yan19}. Sample isochrones ( black dots, see Section \ref{prior})  at solar metallicity ( $\rm [M/H] = 0.0$) are over-plotted on top of MaStar ``good-star" library stars to illustrate our initial guess of the stellar parameters. The ages of isochrones decrease from the bottom to the top along the y-axis. }
\label{fig:gaiacmd}
\end{figure}

We have cross-matched the ‘good spectra’ with Gaia DR2 with a radius of 3.0 arcseconds \citep{Yan19} and found 99\% of the stars with corresponding Gaia magnitudes. The extinction correction was performed by applying \citet{Schlegel98} dust map on the models and comparing the observed spectra with the extincted models to derive the calibration vector.
The various evolutionary phases of the MaStar sample can be inferred from the extinction-corrected color-magnitude diagram shown in Fig.~\ref{fig:gaiacmd}: the main sequence stars are from top left to lower right,
 while the giants are on the top parts, roughly above $M_{G} \leq 2 $. To estimate the values of the stellar parameters, i.e., the Gaia priors, the MaStar sample in the GCMD is matched  with isochrones\footnote{\url{http://stev.oapd.inaf.it/cgi-bin/cmd/}} that have the GCMD information as well as the assigned stellar parameters. We choose a set of isochrones \citep{Bressan12, Chen14, Chen15, Tang14, Marigo17}  with solar metallicity (i.e.,$\rm [M/H]=0.0$) at different ages of 0.003, 0.04, 0.1, 0.2, 0.6, 1, 2, 5 and 12Gyr (this is a wider age range than \citet{Hillmainparam}  as we use the isochrones to constrain the fitting range), and the Kroupa initial mass function (IMF) \citep{Kroupa01,Kroupa02,Kroupa13}.

We set up a first level fitting-subgrid in the space of $ G_{BP} - G_{RP}$ vs. $M_{G}$ as the following: 
$ \delta (G_{BP} - G_{RP}) \leq 1.0 $,  $\delta M_{G} \leq 2.0$, where $\delta (G_{BP} - G_{RP})$ and $\delta M_{G}$ are the differences between a given MaStar target and the isochrones.\footnote{\citet{Hillmainparam} use similar scale of Gaia colors, and a group priors for each star.} The absolute distance  $\sqrt{ (\delta (G_{BP} - G_{RP})^2 + (\delta M_{G})^2}$ is used to estimate the closest prior $\rm T_{eff} $ and $\log g $ from the isochrones for each star ,which we denote as $\rm T_{eff, p}$ and $\log g_{\rm,p}$. 

In the parameter space of BOSZ/MARCS templates, a fitting-subgrid near the assigned prior is set as: 


\begin{alignat*}{2}
\lvert T_{\rm eff,M} - T_{\rm eff,p} \rvert
    &\leq 500\,\mathrm{K},   && (T_{\rm eff,p} \leq 20000\,\mathrm{K}) \\
\lvert T_{\rm eff,M} - T_{\rm eff,p} \rvert
    &\leq 5000\,\mathrm{K},  && (T_{\rm eff,p} \geq 20000\,\mathrm{K}) \\
\lvert \log g_{\rm M} - \log g_{\rm p} \rvert
    &\leq 0.5\,\mathrm{dex}. &&
\end{alignat*}

where $\rm T_{eff, M}$ and $\log g_{\rm,M}$ are the model (BOSZ/MARCS) grid parameters. 
 We chose 500K for most of the stars to cover at least four $\rm T_{eff, M}$ levels in the coarser MARCS grid for stars cooler than 20000K. We chose 5000 K for hot stars to cover as large a range as practical since the total numbers of models with $\rm T_{eff, M}$ larger than 20000K are limited. We use 0.5 dex for $\log g_{\rm,M}$ to cover at least two $\log g_{\rm,M}$ levels in the coarser MARCS grid.  

We do not set the  $ \rm [M/H]$ prior. The final $\rm [M/H]$ is estimated from all the available BOSZ/MARCS models in the $\rm T_{eff}$ and $\log g$ subgrid.
Candidate models with the stellar atmospheric parameters within the box are selected. 
By doing this, the total number of model spectra required to fit a single MaStar spectrum is reduced to a manageable level (typically $ N\sim 1100$ templates), which saves computing time.

\subsection{Estimation of the main stellar parameters $\rm T_{eff}$, $\log g$ and $\rm [M/H]$}

We perform full-spectrum-fitting on the data using ULySS\footnote{\url{http://ulyss.univ-lyon1.fr/}} \citep{Koleva08} with the synthetic stellar spectra mentioned in Section~\ref{sectemp}. 
The algorithm allows the use of Legendre polynomials to account for 
the effect of extinction and/or potential flux calibration systematics.
Normally, a low order (2 to 5) of Legendre polynomials is applied while performing the full-spectrum-fitting. 
The full wavelength range of the MaStar spectra is utilized when fitting the model spectra.
As the line spread function (LSF) varies with wavelength, we convolve the model templates with the varying LSF as a function of wavelength for each specific MaStar spectrum to be processed\footnote{There are two versions of LSF provided by MaStar pipeline. Here we use the one given by the DISP column which assumes the pixel-integration effect is included in the Gaussian broadening kernel. }.
Each fit is characterized by its reduced-$\chi ^2$ value. Detailed extinction calculation is discussed in our paper \citet{Lazarz22}.

The stellar parameter of a given star is estimated by the Bayesian zero assumption, i.e., the uniform prior.
The posterior probability distribution $P \rm (\theta | D)$ of the true
value of the model parameters $\theta$  given observation 
D is proportional to the likelihood function $P \rm (D | \theta) =e^{-\chi^2}$. We assign the probability $P_{i}$ as the weights of individual models.  The expected stellar parameters are expressed as 
\begin{align}
\bar{t} &=
\frac{\displaystyle\sum_{i=1}^{N} P_i t_i}
{\displaystyle\sum_{i=1}^{N} P_i},
\end{align}
where $t_i$ are the model parameters, i.e., $\rm T_{eff, M,i}$, $\log  g_{\rm,M,i}$, 
$[\mathrm{M/H}]_{\rm M,i}$, and $[\alpha/\mathrm{M}]_{\rm M,i}$, $N$ is the total number of models within the fitting-subgrid for each MaStar spectrum. 
The errors in the estimated stellar parameters are determined by their standard deviation

\begin{align}
\epsilon_t &=
\sqrt{
\frac{\displaystyle\sum_{i=1}^{N} P_i
(t_i-\bar{t})^2}
{\displaystyle\sum_{i=1}^{N} P_i}
}.
\end{align}

In order to speed up the calculation process, we first fit for the main stellar parameters  $\rm T_{eff}$, $\log g$ and $\rm [M/H]$ following the above algorithm among models with $[\alpha/M]=0.$
Typically, the derived stellar parameters assigned with the minimum of reduced-$\chi^2$ are close to their initial priors, especially for the prior $\rm T_{eff, p}$. However, there are a few cases where after the first iteration, we find that the minimum reduced-$\chi^2$ is found at the edge of the 
prior subgrid. 
That means that the initial guess of $\rm T_{eff}$ and $\log g$ may not be accurate enough. In this case, we allow the program to take the first iteration result as the new prior 
and run a second iteration to find a better fit for the data.

\subsection{$\alpha$-enhancement parameter $[\alpha/M]$ of the MaStar sample}
Once we obtain the estimates of $\rm T_{eff}$, $\log g$ and $\rm [M/H]$ as described above, we expand the model grids to the dimension of $[\alpha/M]$, 
then repeat the process just outlined
to calculate the $[\alpha/M]$ together with $\rm T_{eff}$, $\log g$ and $\rm [M/H]$ simultaneously. There is no  limit in the $[\alpha/M]$ dimension, and all available $[\alpha/M]$ models that fall into the boxes of $\rm T_{eff}$ and $\log g$ are considered potential solutions. 
The estimated $[\alpha/M]$ is a weight-averaged value within the fitting-subgrid as described above.

\subsection{Compare the methods of different MaStar parameter determination efforts}
Our group has published several sets of stellar parameters for the MaStar project using a variety of methods. This is also documented in our data release {\footnote{\url{https://www.sdss4.org/dr17/mastar/mastar-stellar-parameters/}}}. We describe the differences here to help readers better understand the approach used herein. 

First of all, there are basically two major categories: empirical data training \citep[e.g.,][]{Chenparam20,Imigparam}, and theoretical template spectrum-fitting (e.g.,\citet{Hillmainparam,Hillalphaparam,Lazarz22} and this paper). \citet{Imigparam} derived the stellar parameters through a data-driven algorithm that uses a neural network model, where they promote trustful results from  `data' based on APOGEE stellar parameters calculated from the ASPCAP pipeline. In the stellar parameter regions that are not well represented by ASPCAP, namely higher temperature stars with $\rm T_{eff} >7000 K$, \citet{Imigparam} used theoretical templates \citep[plane-parallel ATLAS9 model atmospheres,][]{Allende18} to extend the derivable parameter space. 

The three analyses that use the theoretical templates of spectrum-fitting are this work, \citet{Hillmainparam,Hillalphaparam} and \citet{Lazarz22}. All use BOSZ templates to derive the stellar parameters of stars, with this work using the finest grid resolution of templates. 
At cool temperatures ($\rm T_{eff, p} < 4500 K$), MARCS templates are introduced in the algorithm of this work (a similar strategy is used in \citet{Hillmainparam}). In the work of \citet{Lazarz22}, only BOSZ templates are utilized.

The differences are also in the fitting methods. \citet{Lazarz22} and \citet{Hillmainparam,Hillalphaparam} both used Markov chain Monte Carlo (MCMC) and interpolated templates. This work uses the fine grid models directly without ever interpolating. 
The \citet{Lazarz22} approach performed fitting 
using multicomponent $\chi^2$ that fit both the individual line features and the broadband continuum with different weights. The work from \citet{Hillmainparam} (using the full spectral fitting code pPXF) and the method presented here (full spectrum fitting) use priors based on the isochrones that match the GCMD. 

We summarize some common features and differences in Table~\ref{tabparamgroup}. A full comparison of our groups various methods of determining parameters and  which perform `better' within specific subsets of parameter space, 
along with recommendations on which parameters to use therein, will be presented in \citet{Yan24}, along with details pertaining to the final DR17 MaStar data release.

\begin{deluxetable}{lrrrllll}

\tablecaption{Summary of the MaStar stellar parameter working group methodologies\label{tabparamgroup}}
\tablewidth{0pt}
\tablehead{
Group & \colhead{Data } & \colhead{BOSZ}    & \colhead{MARCS}  &  
MCMC & \colhead{Interpolation} & \colhead{Gaia } & Broadband \\
name & driven &  &  &  &  & prior & continuum
}
\startdata
Imig et al. & yes &   &   & & yes &  & no \\
Hill et al. & no & yes & yes & yes & yes & yes & no \\
Lazarz et al. & no & yes & no & yes & yes & no & yes \\
Chen et al.   & no & yes & yes & no & no & yes & no\\
(this work)  &   &   &   &   &   &   \\
\enddata
\end{deluxetable}

\section{Validation}\label{secval}

The method used herein is validated by comparing the calculated stellar parameters with the assigned model atmospheric parameters of individual BOSZ models. 
Eleven mock stars based on specific BOSZ models along the H-R diagram with various $\rm [M/H]$ are chosen to assess the overall accuracy of the recovered stellar parameters. 
Here, we limit the models to solar abundance patterns, i.e., $\rm [\alpha /M] =0.0$,  to evaluate the behavior of the main stellar parameters ($ T_{\rm eff}$, $\log g $ and  $\rm [M/H]$).
As mentioned above, the resolution of the MaStar sample is described by the LSF as a function of wavelength. 
The selected BOSZ models are convolved with
the median LSF using Gaussian kernels along the wavelengths axis to create the mock spectra, using the median LSF from MaNGA MPL 9. 
The median error from MaStar MPL 9 is assigned to model the typical MaStar spectral errors: assuming a typical MaStar signal-to-noise ratio per pixel of 100 for the BOSZ spectra, and then propagating with the median MaStar inverse variance. 
The original $T_{\rm eff}$ and $\log g $ values from the BOSZ models are randomly modified to provide slightly different priors, in order to test the accuracy of retrieving the original parameter values upon refitting the perturbed spectra; the adjusted values of $T_{\rm eff}$ range from 0--300K, and those for $\log g $ lie within $+/-0.4$ dex. The metallicity $\rm [M/H]$ remains a free parameter. 
For a detailed validation comparison, the mock stars from BOSZ would need to have the same signal-to-noise ratios as the MaStar sample across different stellar types, which is beyond the scope of this work. We leave this to a forthcoming paper for further investigation.

\begin{figure}
\centering
\includegraphics[scale=0.40,angle=0]{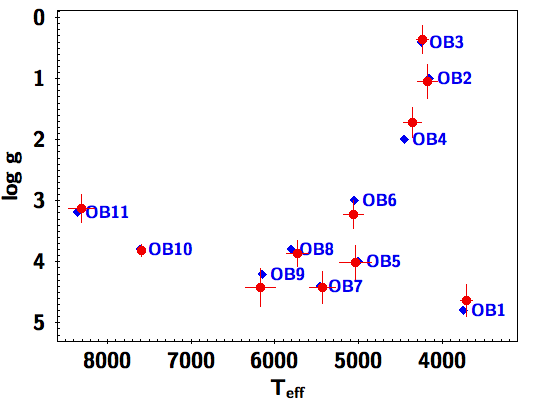}
\includegraphics[scale=0.40,angle=0]{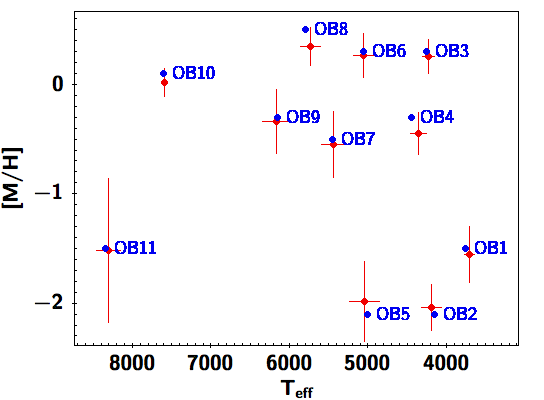}
\caption{Stellar parameters (blue diamonds) of selected BOSZ samples versus their recovered values from this work (red circles). Recovered stellar parameters are consistent with their model parameters within their 1-sigma error bars. The error bars are calculated as the Bayesian standard deviation of all the solutions within the fitting-subgrid.  }
\label{fig:boszhr}
\end{figure}

Figure~\ref{fig:boszhr} shows the recovered stellar parameters for the sample BOSZ stars (red circles) in comparison with their actual values used to create the original models (blue diamonds).
The recovered stellar parameters lie essentially within the 1-sigma uncertainties, thereby indicating the errors associated with deriving stellar parameters using this method. Note that there are no error bars associated with the original BOSZ model values.

\section{Results}\label{secresul}
\subsection{Stellar parameters derived using BOSZ theoretical atmospheric models}\label{secresulmthd}

\begin{figure*}
\centering
   \includegraphics[scale=0.55,angle=0]{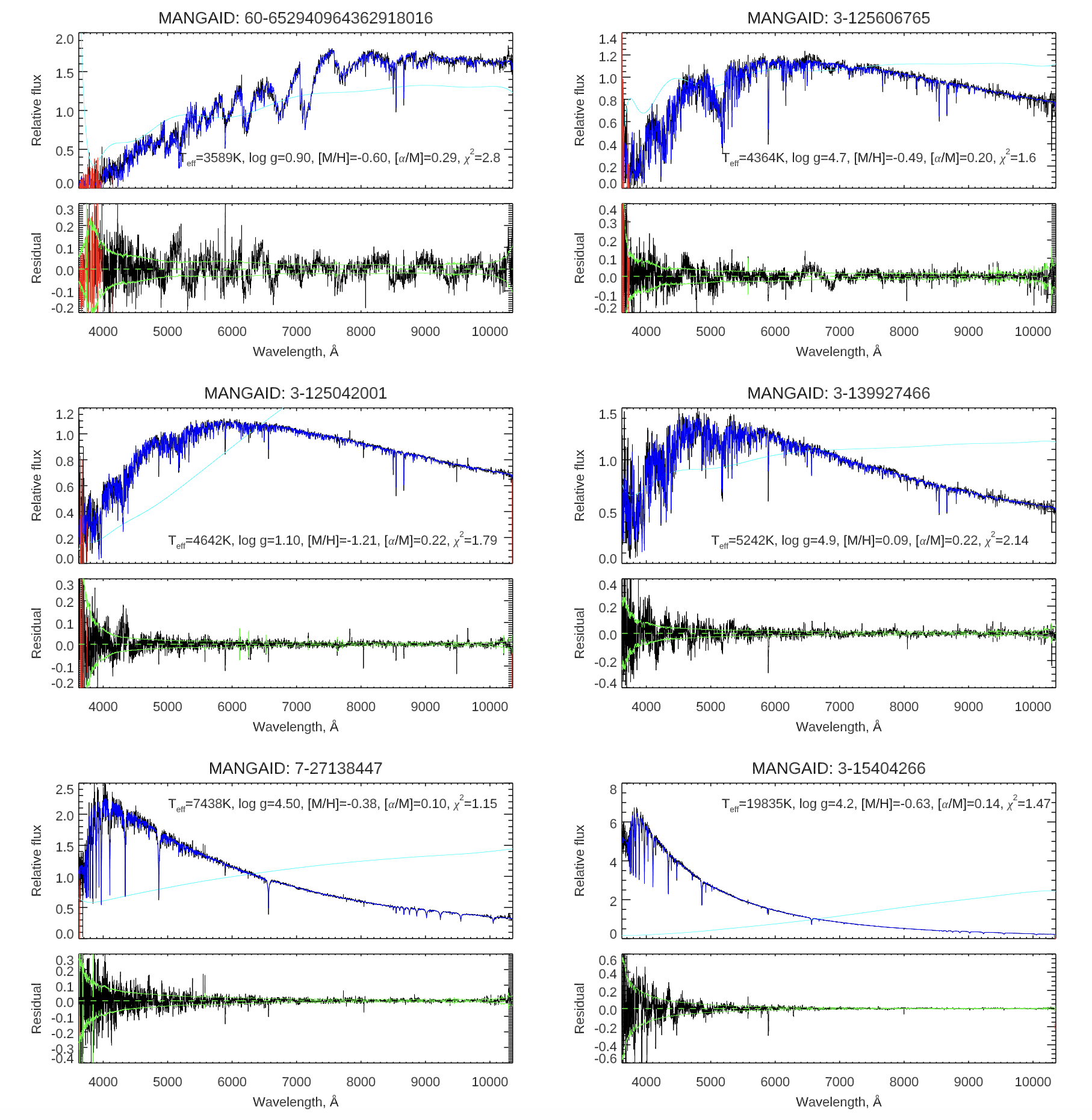}
\caption{BOSZ full-spectrum-fitting examples for the MaStar sample with MANGA ID \text{60-652940964362918016},   \text{3-125606765},  \text{3-125042001},  \text{3-139927466},  \text{7-27138447}, and  \text{3-15404266}. Upper panels: the data is in black, the best-fit is in blue, and the polynomial is in cyan (see Section~\ref{secresulmthd} for details) in each plot. Lower panels: residuals between the data and the best-fit model, with the green line marking 
the one-sigma error from the data. Red pixels mark the bad pixels flagged in the data.  }
\label{fig:boszfullfit}
\end{figure*}

We have successfully derived stellar parameters for 56,485 spectra (96\% of the total sample of MaStar good spectra){\footnote{The failed fits are mostly due to template mismatches or lie outside the limited stellar parameter range of the theoretical spectral models.} }
 using BOSZ templates. The stellar parameters using BOSZ templates cover a range of effective temperature $\rm 3500 \leq T_{eff} \leq 29861 K$,  
of surface gravity $0 \leq \log g \leq 5.43$, of metallicity $-2.5 \leq \rm[M/H] \leq 0.5$, and of $\alpha$-enhancement $-0.25 \leq \rm [\alpha /M] \leq 0.5$.  
Some examples of the full-spectrum fits to MaStar spectra using BOSZ templates is shown in Fig.~\ref{fig:boszfullfit}. The observed MaStar spectra are marked in black lines, and the best fits from BOSZ theoretical templates are marked in blue lines. The polynomials used to correct for differences in the broadband shape due to extinction effects and/or flux-calibration residuals are marked in cyan lines. The bad pixels flagged by MaStar's mask flag are marked in red lines. The lower panel shows the residual between the data and the model, with the green spectrum marking the one-sigma error level. The calculated stellar atmospheric parameters and reduced-$\chi^2$ are provided in each of the panels.

\begin{figure}
\centering
\includegraphics[scale=0.55,angle=0]{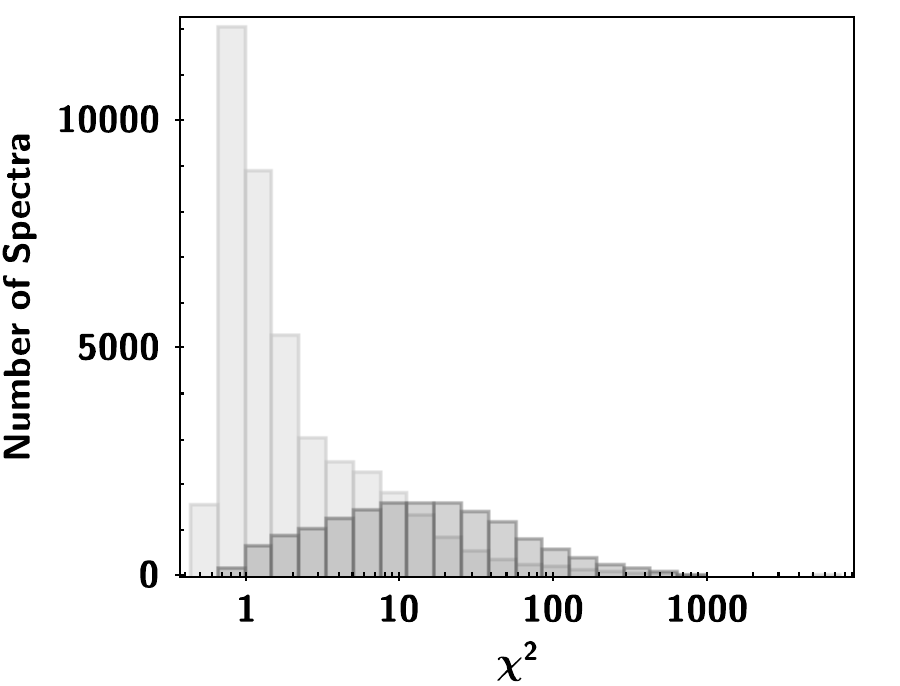}
\caption{Distribution of $\chi^2$  from BOSZ-based parameters for warm ($T_{\rm eff} > 5500 K$, light grey) stars and cool ($T_{\rm eff} < 5500 K$, dark grey) stars separately. }
\label{fig:chi-bosz}
\end{figure}

The resulting reduced-$\chi^2$ distribution using the BOSZ template 
is shown in Fig.~\ref{fig:chi-bosz}. 
There are basically two groups of stars, as shown in 
Fig.~\ref{fig:chi-bosz}: warm stars with $\rm T_{eff} \geq 5500 K$ forming the first peak of $\chi^2\sim1$, and cooler stars with $\rm T_{eff} \leq 5500 K$ forming the second peak of $\chi^2\sim 20$. Fits with larger values of reduced-$\chi^2$ are likely due to the unsuccessful template match, especially at the cool star ranges. 
This is because theoretical models are challenged to reproduce cool molecular bands \citep[e.g.,][]{Conroy13}. %
MARCS models were specifically constructed to help address the deficiencies in the physics at cool temperatures, particularly with regard to molecules. However, as the MARCS models also deviate from the observed spectra in  certain spectral regions, further improvements in the underlying physics of cool stars are likely still needed.

\subsection{Stellar parameters for cool stars derived using MARCS theoretical atmospheric models}

\begin{figure}
\centering
\includegraphics[scale=0.55,angle=0]{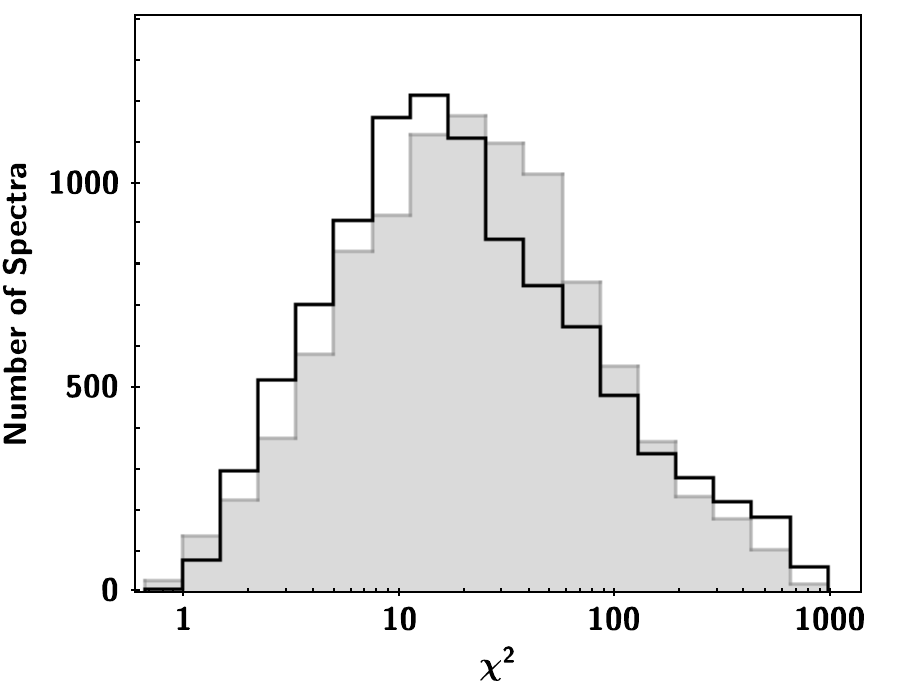}
\caption{Distribution of $\chi^2$ of stars with $T_{\rm eff} < 5500 K$ fitted with both MARCS and BOSZ models. The MARCS -based parameters (open black line) with $2600 K <T_{\rm eff} < 5500 K$ is overplotted on top of the BOSZ results (light grey) with $\rm T_{eff} < 5500 K$. $\rm T_{eff} \leq 5500\ K$ is chosen for the cool end of parameter space in order to explore the best solutions from theoretical atmospheric results.}
\label{fig:chi-marcs}
\end{figure}

We identify cool stars in the MaStar sample by selecting on the $BP-RP$ color. 
Isochrone models with  $BP-RP \geq 1.3$ have $T_{\rm eff, p} \lesssim 4500K$.
Assuming that our stars are accurately matched with their Gaia counterparts, their $BP-RP$ color is therefore a strong indicator of their temperature scales. We have selected 10,519 cool stars ($\sim 18\%$ of the full sample of	 MaStar) to analyze their stellar parameters with the MARCS templates, in addition to BOSZ templates.. 
The process of estimating stellar parameters is the same as using the BOSZ templates. The only difference here is the template grid size, i.e., the MARCS template grids have larger $T_{\rm eff}$ and $\log g$ steps compared to the BOSZ grids. 
We have determined $\sim17\%$ (10,030 spectra) stellar parameters of the entire 59,266 MaStar spectra sample using MARCS templates. The reduced-$\chi^2$ distribution from MARCS models is shown in Fig.~\ref{fig:chi-marcs}. The distribution of $\chi^2$ from BOSZ-based cool stars is presented for comparison (light grey). The peak of the distribution is around $\chi^2 ~\sim 20$ from both set of templates, where MARCS models show the peak at a slight smaller $\chi^2 \sim 15$ value than the BOSZ models. As expected, at cooler temperatures, the fits are not as accurate as in the bench-mark stars ($T_{\rm eff} \sim 4000 - 6500 K$), partly due to the model quality. Certainly, there is  some room to improve the cool star atmospheric models. 

\subsection{Combining the results from two sets of templates for the cool stars} \label{comb2mods}
\begin{figure} 
\centering
\includegraphics[scale=0.55,angle=0]{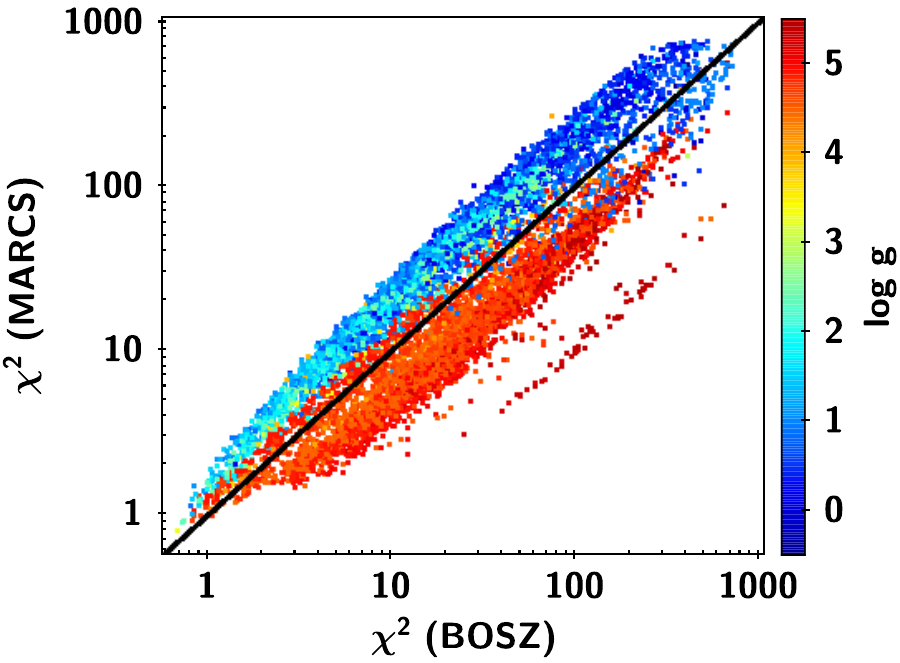}
\caption{Reduced-$\chi^2$ distribution from BOSZ templates compared with MARCS templates in full-spectrum-fitting 
for stars with $\rm T_{eff} \lesssim 5500K$. 
The color bar indicates the adopted $\log g$, where dwarfs are in red, and giants are in blue.  A black line representing the one-to-one relationship is presented to guide the perfect correlation.}
\label{fig:comchi2logg}
\end{figure}

\begin{figure}  
\centering
\includegraphics[scale=0.55,angle=0]{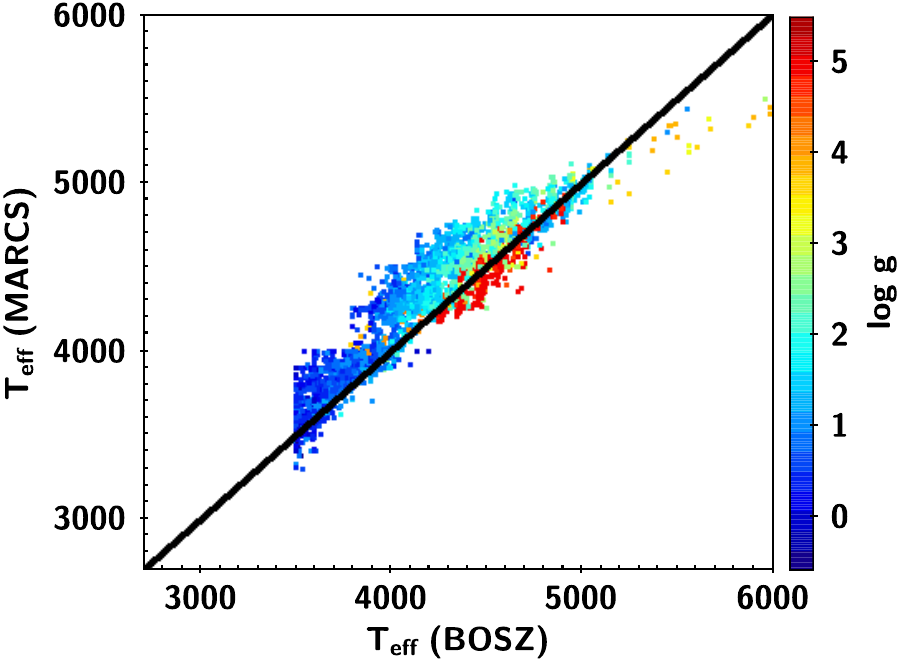}
\includegraphics[scale=0.55,angle=0]{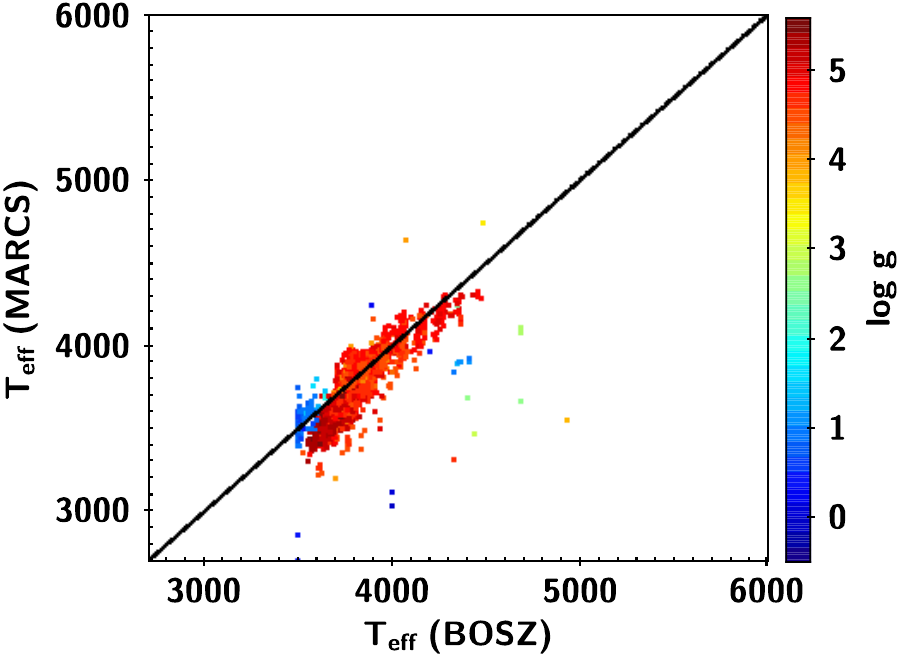}
\caption{Best-fit $T_{\rm eff}$ values for stars with final $T_{\rm eff} \lesssim 5500K$ that have 
been fit by both BOSZ and MARCS templates. 
Left panel shows those stars whose final adopted stellar parameters are derived based on BOSZ templates. Right panel shows those stars whose final adopted parameters are derived based on MARCS templates. 
The color bar indicates the adopted $\log g$, where dwarfs are in red and giants are in blue.  A black line representing the one-to-one relationship is presented as a guide.} 
\label{fig:comteffogg}
\end{figure}

A total of 10162 visit spectra of cool stars have their estimated stellar parameters with both MARCS and BOSZ template-fitting. 
We check the reduced-$\chi^2$ for each star from both MARCS and BOSZ fits.{\footnote{Note that the denser BOSZ grid may allow for lower $\chi^2$ to some extent.}} The distribution of the reduced-$\chi^2$ is presented in Fig.~\ref{fig:comchi2logg}, with the adopted $\log g$ shown by the color coding. The set of parameters with smaller reduced-$\chi^2$ are adopted \citep[as in][]{Hillmainparam,Hillalphaparam} as our final stellar parameters for that certain star. A clear bias between MARCS and BOSZ fits shows that cool dwarf stars prefer models from MARCS rather than the BOSZ templates. However, the giants and subgiants in this temperature range (4500K$\lesssim \rm T_{\rm eff} \lesssim 6500K$) are better modeled by the BOSZ templates.  This bias is clearly presented in Fig.~\ref{fig:comteffogg}, where we show the stellar parameter-distribution, especially $T_{\rm eff}$, between the BOSZ and MARCS models. 

\begin{figure*}[ht!]
\centering
   \includegraphics[scale=0.51,angle=0]{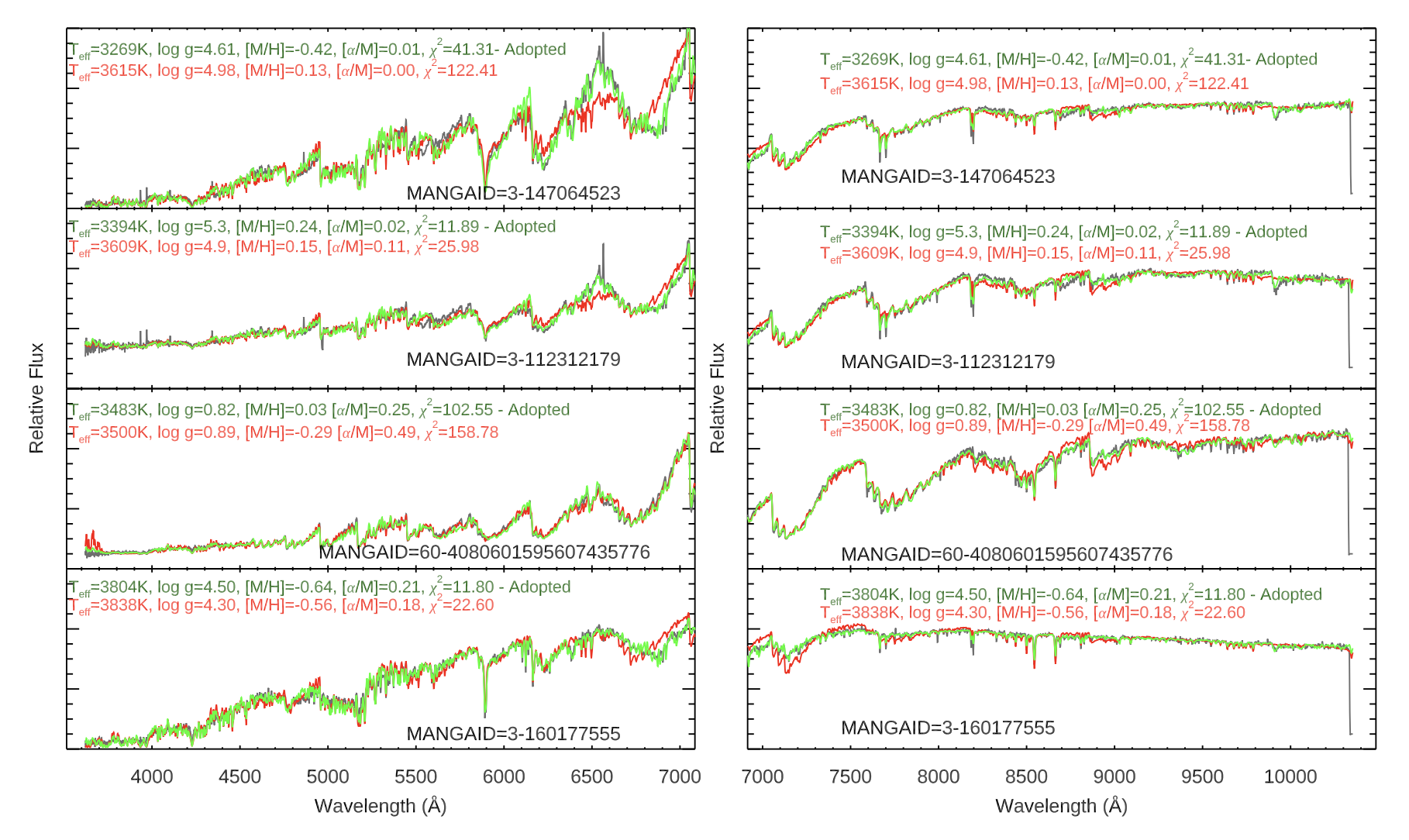}
  \caption{Examples of MARCS models provide the better full-spectrum-fitting 
 for the cool end of the MaStar sample:
 MANGA ID \text{3-147064523},   \text{3-112312179},   \text{60-4080601595607435776}, and \text{3-160177555}. 
Data spectra are in dark gray,  with their best fits from both BOSZ (red) and MARCS (green). 
Adopted stellar parameters are shown in black text. Stellar parameters from BOSZ fitting are in red text for comparison.
In general the MARCS models fit better, especially for the low-$\rm T_{ eff}$ dwarfs.}
\label{fig:marcsbbosz1}
\end{figure*}

\begin{figure*}[ht!]
\centering
   \includegraphics[scale=0.51,angle=0]{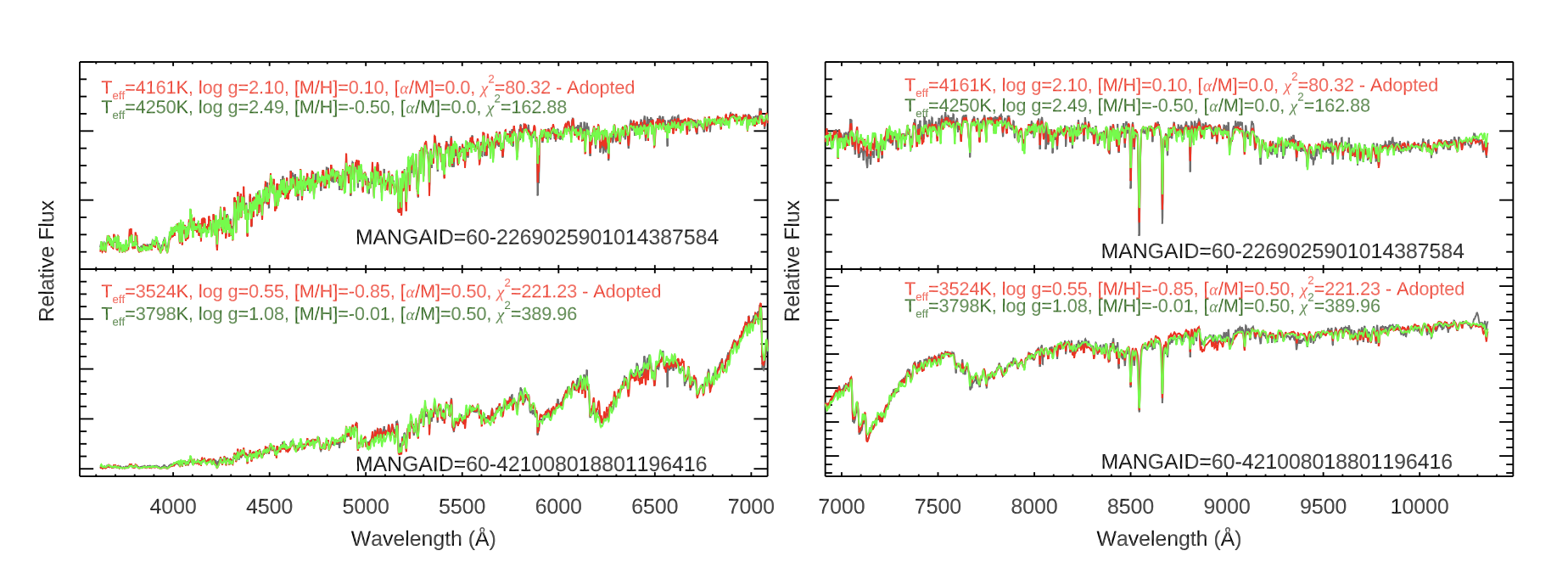}

 \caption{Two examples in which BOSZ has better fitting results than MARCS, symbols are the same as Fig.~\ref{fig:marcsbbosz1}:   MANGA ID \text{60-2269025901014387584} and MANGA ID \text{60-421008018801196416}. Adopted stellar parameters are shown in black text. Stellar parameters from MARCS fitting are in green text for comparison. For the majority of the cool giants the BOSZ models provide better fits.
}
\label{fig:marcsbbosz2}
\end{figure*}

We present some full-spectrum-fitting examples with both templates in Fig.~\ref{fig:marcsbbosz1} and Fig.~\ref{fig:marcsbbosz2}, with the adopted stellar parameters listed in each panel. Figure~\ref{fig:marcsbbosz1}  show examples where MARCS has better best-fits than BOSZ, while the Fig.~\ref{fig:marcsbbosz2} shows the opposite. 
As mentioned above, the BOSZ templates have a minimum $T_{\rm eff}$ boundary of 3500K (their lowest temperature limit), which may not provide an adequate fit to all line depths. When the actual observation is cooler than the model temperature range the fitting program can only choose the closest features, not being able to recover the line features. This is the case for MAGAID \text{3-147064523}, \text{3-112312179}, and \text{60-4080601595607435776}.  

\begin{figure}
\centering
\includegraphics[scale=0.54,angle=0]{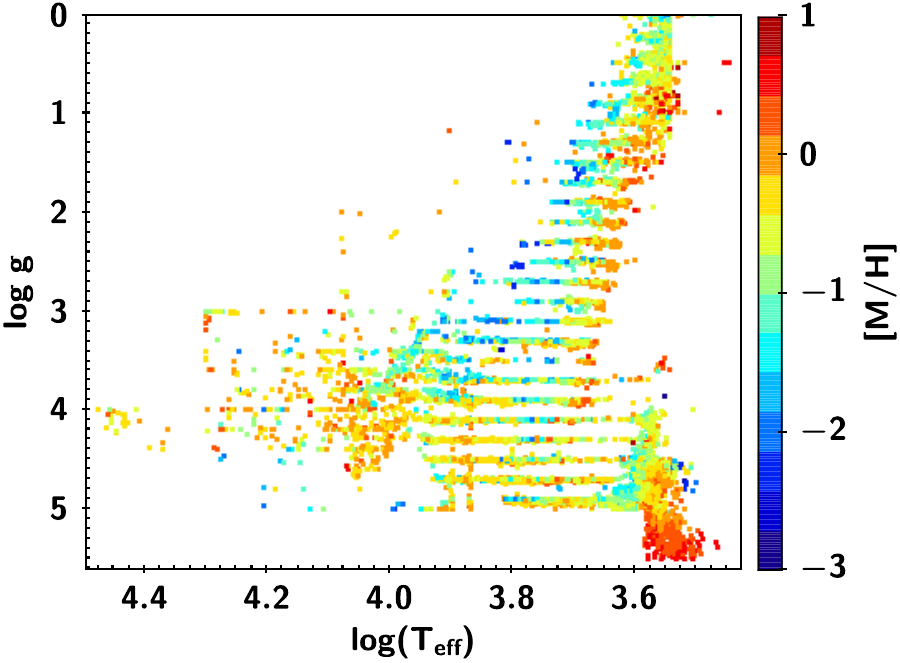}
\includegraphics[scale=0.54,angle=0]{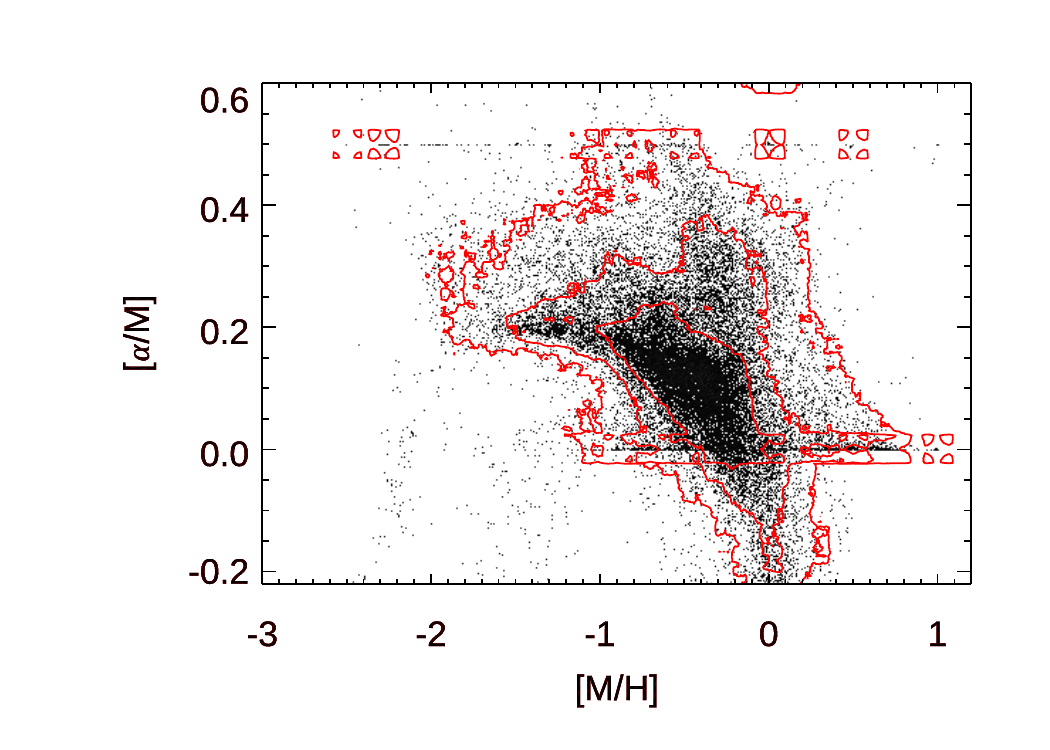}
\caption{Stellar parameters from this work. Left panel shows the $\log g $ vs. $T_{\rm eff}$, with metallicity  $\rm [M/H]$  color-coded on the side bar. Right panel shows the adopted $[\alpha/M]$ as a function of  $\rm [M/H]$ from this work. The red contours are generated using IDL code \texttt{density.pro}, with three contour levels marking 0.04, 0.2, and 0.5. Parameters are cleaned by applying an uncertainty threshold of  $\rm [\alpha/M]$ less than 0.27 dex (see Section~\ref{alphaqc} for details). } 
\label{fig:paramscomb}
\end{figure}

The full set of adopted stellar parameters, after combining the MARCS and BOSZ results, is presented in Fig.~\ref{fig:paramscomb}.  In the left panel, we show the overall stellar parameter distributions of the three main parameters, namely $T_{\rm eff}$, $\log g $, and $\rm [M/H]$. The adopted stellar parameters per visit show a well-populated parameter space from cool dwarfs to the hot end. In the right panel, we show $\rm [M/H]$ vs. $[\alpha/M]$.  All visit spectra which have been flagged due to issues have not been included in this plot (see Section~\ref{alphaqc} for details). The red contours mark the density levels of 0.04, 0.2, and 0.5. 

\subsection{Standard stars for flux calibration}

\begin{figure}
\centering
\includegraphics[scale=0.55,angle=0]{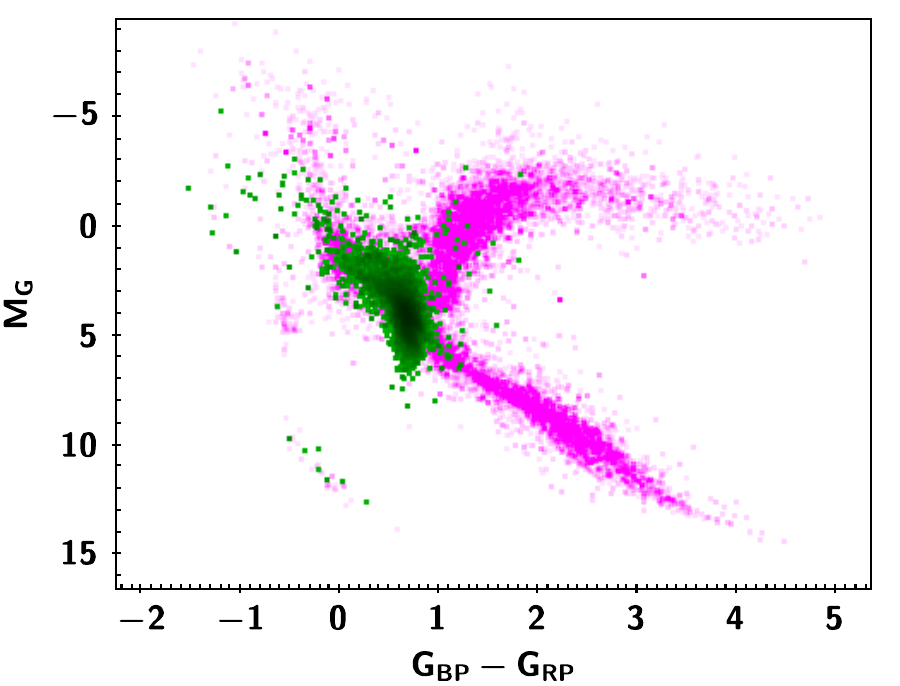}
\includegraphics[scale=0.55,angle=0]{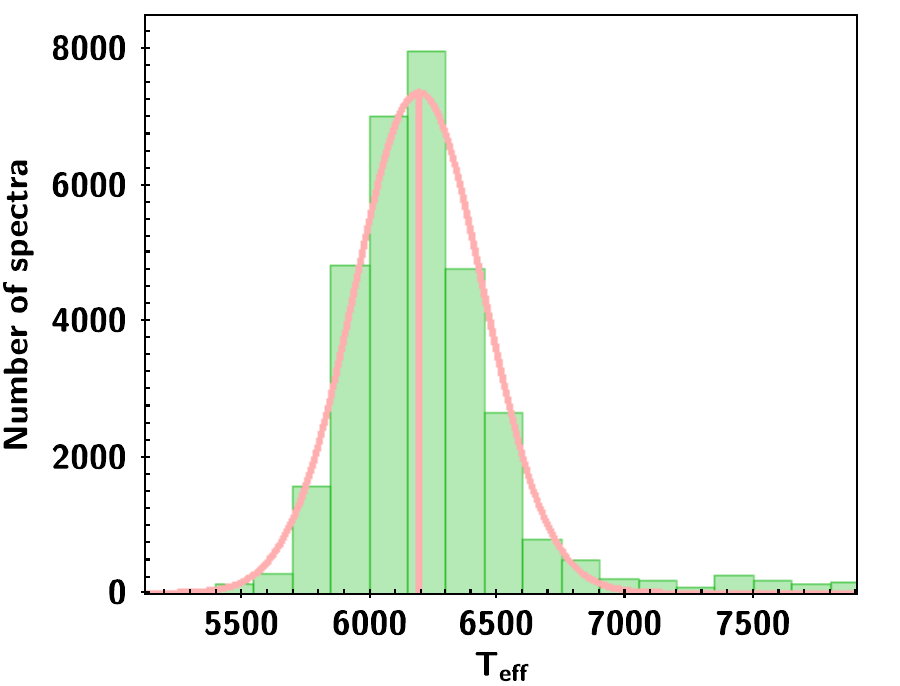}
\caption{Left panel shows the flux standard stars (green dots) in the MaStar sample presented in the GCMD diagram. The white dwarfs were intentionally targeted to provide better flux calibrations. The right panel shows the distribution of adopted effective temperatures calculated in this work for the flux standard stars (i.e., the green dots on the left panel). Most flux standard stars are in the range of $5500K \lesssim T_{\rm eff} \lesssim 7000K$, with the Gaussian peaked at 6190K. The standard deviation of the Gaussian is 250K for the flux standard stars.  }
\label{fig:stdstar}
\end{figure}

In building the MaStar stellar library, we have selected flux standard stars (primarily late-F type main-sequence stars) by their photometry criteria \citep{Yan16,Yan19}. In brief, the standard stars are selected based on observed magnitudes between 14.5 and 17.2 in the $g$-band for some fields, while the majority of standard stars are selected by the SDSS color-color diagram, where the SDSS colors are converted from known PS1 and APASS literature values \citep{Yan19}.
Users can find the flux standard stars with the identification of ``7" in  ``IFUDESIGN=7xx''. The numbers run from 701 to 712. We show them in the GCMD in Fig.~\ref{fig:stdstar}. These flux standard stars have an over-dense region at $\rm (G_{BP}-G_{RP}) \sim 0.7$ and $\rm M_{G} \sim 4.3$. Most of the standard stars, as 
expected, have $T_{\rm eff} \sim 6000$K, including the white dwarfs used as flux standards. However, there are about 1400 stars with higher effective temperatures, whose $T_{\rm eff} $ extends to 28,500 K. The vast majority ($99.4\%$) of these standard stars have $\log g \geq 3.5$.

\section{Quality control}\label{secqc}

We fitted 59,266 observations for 24,281 unique stars in MaStar. In order to make sure the stellar parameters derived from 
the full-spectrum-fitting process are sensible, validation of the spectral fitting on a case-by-case basis is needed.

The BOSZ/MARCS model grids have limited parameter coverage. The initial guess taken from 
the isochrones is likely to extend beyond the grid boundaries of the theoretical models, especially when the stars in the GCMD are too cold or too hot, will render invalid results and those cases are not included 
in this work (stellar parameters are set to ``-999" for these stars). Moreover, the algorithm does not return valid stellar parameters in a few cases due to  mismatched templates; 
these stellar parameters are also set to `-999'. 
This results in $\sim4\%$  of the good stars having no associated stellar parameters.

\subsection{Best fit assessment}


Here we are assessing how consistent the prior is with the spectra-fitting result. The inconsistency can come from multiple sources. The contributors to potential inconsistency include (1) discrepancy between the isochrone and BOSZ in their CMD prediction, (2) incorrect galactic extinction correction from 3D dust map, (3) metallicity dependence of isochrone, (4) bad spectra fitting.
The deviation in $\rm T_{eff}$, defined as $\Delta\rm T_{eff} = \lvert T_{eff, p} - T_{eff} \rvert$, being the major prior assumption, is monitored in the quality control process. A standard deviation of 300K is observed among all the $\Delta\rm T_{eff}$. We filtered out the stars with $\Delta\rm T_{eff}$ larger than twice the fitting-subgrid in the dimension of effective temperatures, whose stellar parameters are considered unreliable due to inconsistency. Those stars may suffer from template grid limitations, or inaccurate GCMD {\it Gaia} information (inconsistency of $G_{BP} - G_{RP}$ color vs. isochrone prediction is observed in Fig.~\ref{fig:gaiacmd} at  $G_{BP} - G_{RP} \geq 3$ for the giants). A detailed discussion of the GCMD assignment is beyond the scope of this paper and will be addressed in future work. 


We also use a threshold in fractional deviation of $\rm T_{eff}$ ($\Delta\rm T_{eff} / T_{eff}$)  
to filter out template mismatches, especially at lower temperatures where the atmospheric models suffer from accurately reproducing molecular features. 
To do this, we adopt $\Delta\rm T_{eff} / T_{eff} = 0.12$ for $\rm T_{eff} \leq 9000K$. We adopt $\Delta\rm T_{eff} / T_{eff} = 0.06$ for $\rm T_{eff} \geq 20000K$ to optimize the template matches at high-temperature stars. 
We combine the above  $\Delta\rm T_{eff}$ requirements and the $\Delta\rm T_{eff} / T_{eff}$ ratio to identify all potential inconsistencies due to any of the above reasons.

 In addition to the above issues, we note that some stars have their uncertainty of stellar parameters returned as an unrealistically small number, close to zero. This occurs when the difference in $\chi^2$ between adjacent grid points is sufficiently large, 
resulting in a high probability of one grid point dominating over the others.
As a result, the final parameters are on the grid points, and the uncertainty returned is nearly zero. The number of such cases can be reduced with 
a denser model grid in order to get more sensible fitting uncertainties and refined parameters. 

We note that there are some ``quantized" $\log g$ values in Fig.~\ref{fig:paramscomb}. This mostly occurs in stars with temperature $\rm T_{eff} \leq 9000K$. We usually find two candidate $\log g$ values next to each other satisfying the fitting procedure. When averaging them with Bayesian method, the derived $\log g$ are in the middle of the grids. The artifact on the right panel of Fig.~\ref{fig:paramscomb} at $\rm [\alpha/M]=0$ are likely from the effect that we have finer solar models with steps of $\rm T_{eff}=$50K, but less dense non-solar models, i.e., $\rm [\alpha/M] \neq 0$ with steps of  $\rm T_{eff}=$100K. Such cases could be improved by involving finer non-solar models.

Another issue is that some stars in our sample could be variable stars. We are using the epoch-averaged Bp-Rp color and the absolute magnitude derived from the epoch-averaged G magnitude. For variable stars, we expect this to match well with the isochrone models. 
Our per-visit spectra, they are not necessarily identical with the average spectra and may
lead us to derive different stellar parameters for the individual visits.
This could lead to large $\Delta T_{\rm eff}$ or $\Delta \log g$, which could lead to rejection of some epochs. We leave the identification all variable stars in the catalog and the study of their parameters
for future work.


\subsection{$\rm [\alpha/M]$ parameter assessment}\label{alphaqc}

While calculating $\rm [\alpha/M]$, we observed some low-quality fits, particularly among the flux standard stars featured in the GCMD plot (see Fig.~\ref{fig:stdstar}).
We note that this is not an issue with the BOSZ templates 
but rather due to the observed spectra having little distinguishing power for $\rm [\alpha/M]$. Most of these flux standard stars are metal-poor F-stars. Consequently, they appear fainter and have larger uncertainties in their Gaia G-band magnitude. Given the flat prior and the Bayesian averaging process of our approach (within the fitting-subgrid), the final $\rm [\alpha/M]$  values are pulled toward the simple averages of several discrete grid points. As a result, the stellar parameters of these stars exhibit a degeneracy between $\rm  [M/H]$ and $\rm [\alpha/M]$.
We adopt an upper threshold of 0.27 dex for the uncertainty in $\rm [\alpha/M]$ to flag an $\rm [\alpha/M]$ measurement as invalid. A similar quality control cut is applied in \citet{Hillalphaparam}. This criterion flags $43\%$ of the visit spectra from the
total number of MaStar observations. The cleaned distribution of $\rm [\alpha/M]$ is presented in the right panel of Fig.~\ref{fig:paramscomb}. However, when users are primarily interested in the main stellar parameters, i.e., $\rm T_{eff}$, $\log g$ and [M/H], the $\rm [\alpha/M]$ flag should not be used as a criterion to filter out poor fits. 

As mentioned above, the formal uncertainties of some stars are reported as zero. In these cases we adopt
half of the grid size as the uncertainties.

\section{Consistency}\label{secconsis}  
\subsection{Internal consistency}

The internal parameter consistency is assessed by examining 
repeated observations of a given star. Normally, each MaNGAID corresponds to one unique star, except for a few cases, which are described in the MaStar data Caveats section {\url{https://www.sdss4.org/dr17/mastar/mastar-caveats/}} (see section `Stars with more than one MaNGA-IDs'). The total number of unique MaNGAID for the 59,266 spectra is 24,290, of which 13,360 have repeated observations. We evaluate the parameter distribution of repeated observations for each star. Typically, 2  to 4 visits (repetitions, same plate number with different MJDs) belong to the same star. 
We measure the ``effective standard deviation" by dividing the median absolute deviation among multiple visits to the same star by 0.6745 \citep{Beers1990}.

\begin{figure}
\centering
\includegraphics[scale=0.8,angle=0]{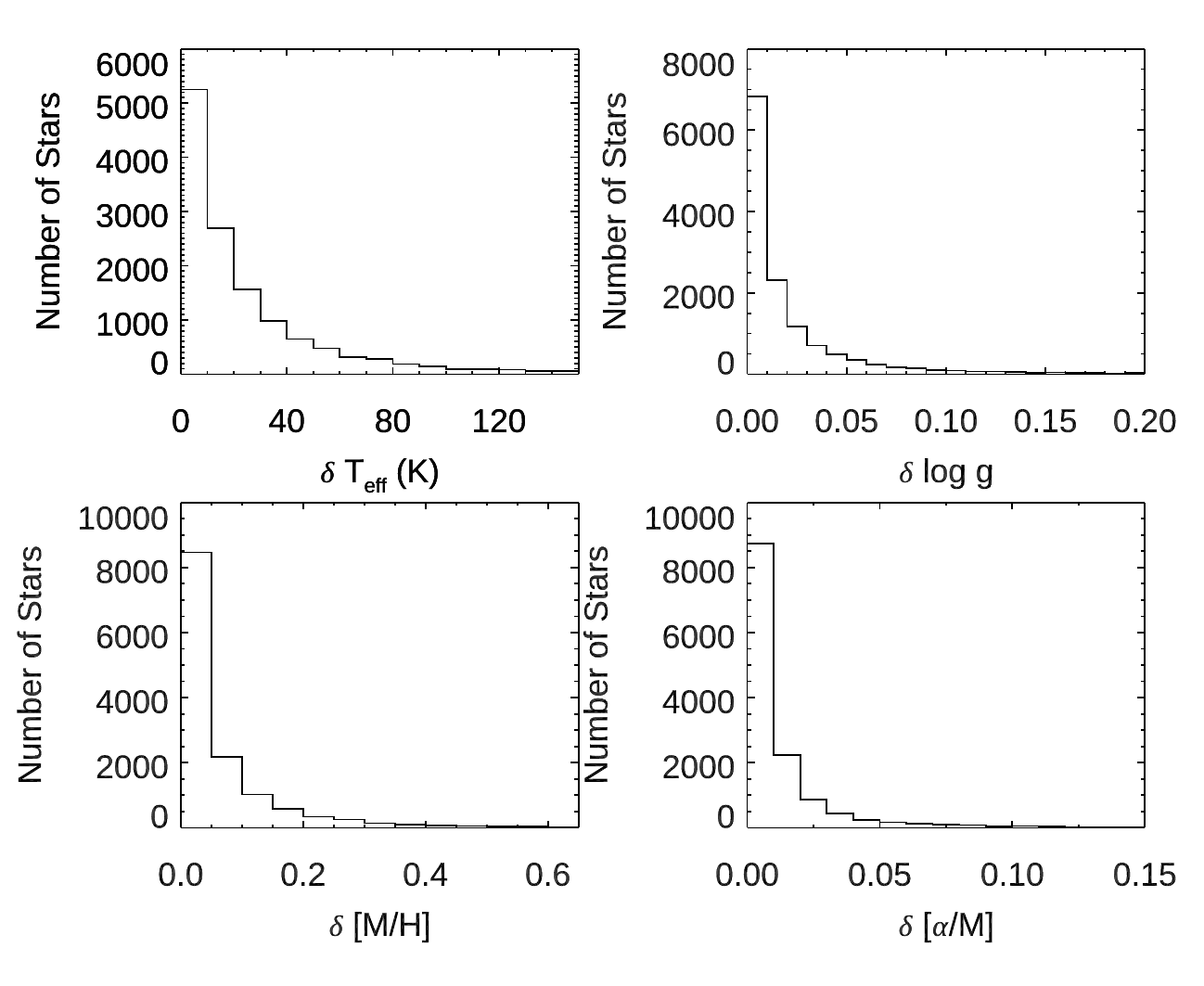}
\caption{Distribution of the effective standard deviation (see text for details) of stellar parameters for repeated observations from 13360 unique MANGAIDs. }
\label{fig:stdintrparam}
\end{figure}

The histogram of the effective standard deviations for all four parameters is shown in Fig.~\ref{fig:stdintrparam}. 
Among all the stars with repeated observations, 90\% have an effective standard deviation of $T_{\rm eff} $ less than 75 K. The 90th-percentile of the effective standard deviation is 0.175 dex for $\rm [M/H]$, 
0.065 dex for $\rm \log g$, and 0.035 for $\rm [\alpha/M]$. 


\subsection{External consistency}



\begin{figure}[h]
\centering
\includegraphics[scale=0.70,angle=0]
    {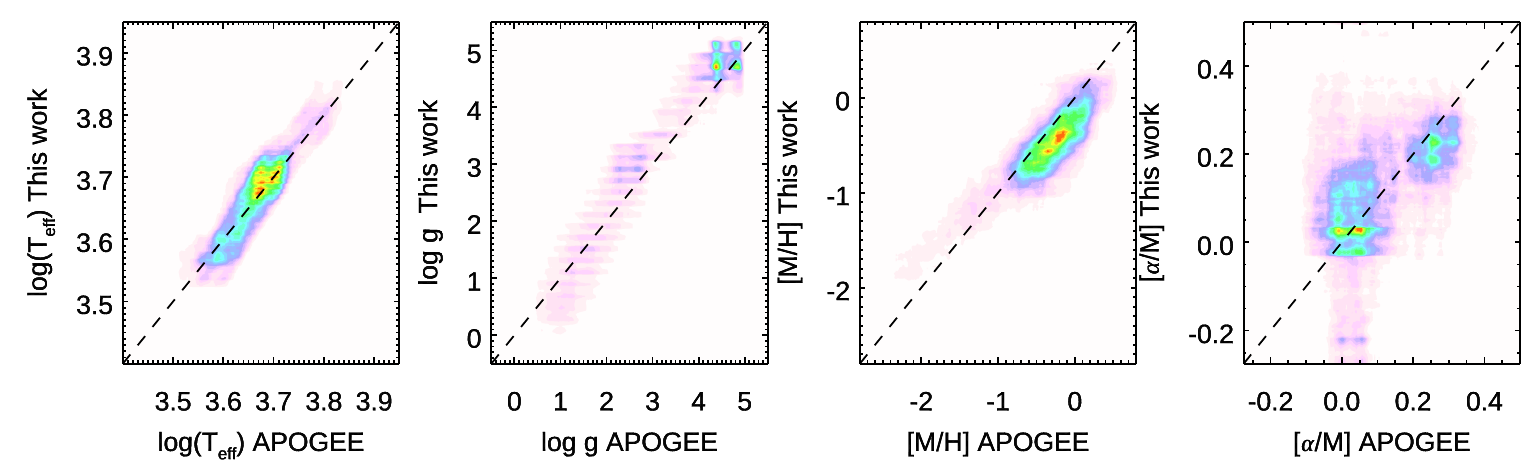}
\caption{The stellar parameters derived from this work were compared with the stellar parameters from APOGEE.
A density map is generated in each panel to show the distribution between this work and the APOGEE results, where the red color corresponds to the highest density. The black-dashed lines mark the one-to-one relationship to guide the perfect correlation. }
\label{fig:para2apo} 
\end{figure}

The MaStar collection shares a significant number of stars with other stellar libraries and surveys. 
The APOGEE stellar parameters are derived with the ASPCAP \citep{Gacia16, Holtzman18, Jonsson20} pipeline, which automatically analyzes the high-resolution ($R\sim22,500$) spectra in the H-band using interpolated ATLAS9 synthetic atmospheric models. We show the stellar parameters in common between APOGEE and the MaStar parameters derived herein in Fig.~\ref{fig:para2apo}. The effective temperature between APOGEE and MaStar shows excellent agreement, with a typical standard deviation of 200K. A slight systematic shift in $T_{\rm eff}$ is observed, namely, our $T_{\rm eff} $ is $\rm \sim 60K$ warmer than the APOGEE result. This is more evident for the stars above $\rm T_{\rm eff} \geq 4500K$ (i.e., the dense clump near $\log T_{\rm eff} \sim 3.66$ in Fig.~\ref{fig:para2apo}). 
For stars that are around 4000K, the APOGEE result is promoting a slightly warmer  $T_{\rm eff} $. The slight systematic difference in  $T_{\rm eff}$ is likely due to the templates, i.e., BOSZ and MARCS synthesis atmospheric models. 
We note that APOGEE focuses on lower temperature stars \citep{Lazarz22}, and the comparison of $T_{\rm eff} $ is dominated by $\rm 3500K \leq T_{\rm eff} \leq 8000K$ ($\sim 90\%$ of all common matches).  For higher-temperature stars, we need to find other resource(s). 

A comparison of the $\rm \log g$ between this work and APOGEE shows good general consistency, with a higher mean value from our result by $\sim 0.16$ dex and a standard deviation of 0.45 dex. There is, however, a clear deviation around $\rm \log g \sim 2.5$, which roughly divides the group of giant stars ($\rm \log g \leq 2.5$) from the dwarfs ($\rm \log g \geq 2.5$). Looking into the details of the giants shows an excellent agreement of $\rm \log g$, with a tiny mean shift of 0.03 dex (standard deviation of 0.53 dex). The systematic shift for the dwarfs is larger, with about 0.26 dex (a standard deviation of 0.34 dex) higher $\rm \log g$ than APOGEE. 

The metallicity $\rm [M/H]$ in the third panel of Fig.~\ref{fig:para2apo} shows a systematic bias towards the metal-rich ($\rm [M/H] \geq -1.0$) group, where we find lower values of the metallicity by $\sim$ 0.17 dex.  
For the metal-poor stars, we seem to find larger values of $\rm [M/H]$ by about 0.22 dex. The overall metallicity $\rm [M/H]$ is consistent within 0.3 dex rms.  

The  $\rm [\alpha/M]$ distribution in the last panel of Fig.~\ref{fig:para2apo} shows consistency with a wider scatter. 
The RMS scatter in Y-direction is 0.16 dex, which is consistent with the combination of the median uncertainty of our  $\rm [\alpha/M]$ measurements (0.18 dex) and that of APOGEE's measurements (0.07 dex).
It is clearly seen that there are two populations, i.e.,  $\rm [\alpha/M] \leq 0.15$ (solar-scaled stars) and $\rm [\alpha/M] \geq 0.15$ ($\alpha$-enhanced stars). These two groups, although not clearly distinguishable by  their metallicity ($\rm [M/H]$), have different metallicity distributions. Namely, 
the stars having roughly solar  $\rm [\alpha/M]$ have metallicities ranging from metal-rich to solar ($\rm [M/H] \geq -0.7$) 
and the $\alpha-$enhanced stars are mostly dominated by lower metallicities ($\rm [M/H] \leq -0.7$). \textit{Dividing the metal-rich to solar metallicity from the lower metallicity with $\rm [M/H] \sim -0.7$ is neither optimal nor consistent with the above metallicity-consistency discussion,  but rather a guideline here. }

We note that by comparing with APOGEE, our  $\rm [M/H]$  seems to have a systematic shift towards the metal-poor direction. Moreover, a slight curvature is observed in the panel of  $\rm [M/H]$ comparisons in Fig.~\ref{fig:para2apo}.  We offer the conversion between our metallicity $\rm [M/H]_{mastar}$ and the APOGEE values $\rm  [M/H]_{ToAPO}$ using the following formula:

\begin{equation}
{\rm [M/H]_{ToAPO} =-0.2543 \times {\rm [M/H]_{MaStar}} ^2 +0.8007 \times {\rm [M/H]_{MaStar}} +0.1329 }  
\end{equation}
The RMS of {$\rm [M/H]$} residual relative to the APOGEE ASPCAP values is 0.20 dex.

As mentioned above, our estimations of stellar parameters are based on the Gaia color--magnitude relation as priors. It makes sense to check the parameter consistency with the Gaia team results when possible.  
Recently, the Gaia DR3 data release \citep{gaiadr3} published stellar parameters produced by the Astrophysical parameters inference system (Apsis). The technical details of Gaia stellar parameters are described by \cite{Creeveygaia22} and \cite{Fouesneaugaia22}.  
In this work, we compare Gaia DR3 parameters under the General Stellar Parametrizer from Photometry (GSP-Phot) category, and The General Stellar Parametrizer from spectroscopy (GSP-Spec). The GSP-Phot provides the effective temperature,  logarithm of surface gravity, metallicity [M/H], magnitude, and extinction information, while the GSP-Spec provides all four stellar parameters as in our work plus individual chemical abundances.

\begin{figure}
\centering
\includegraphics[scale=0.70,angle=0] {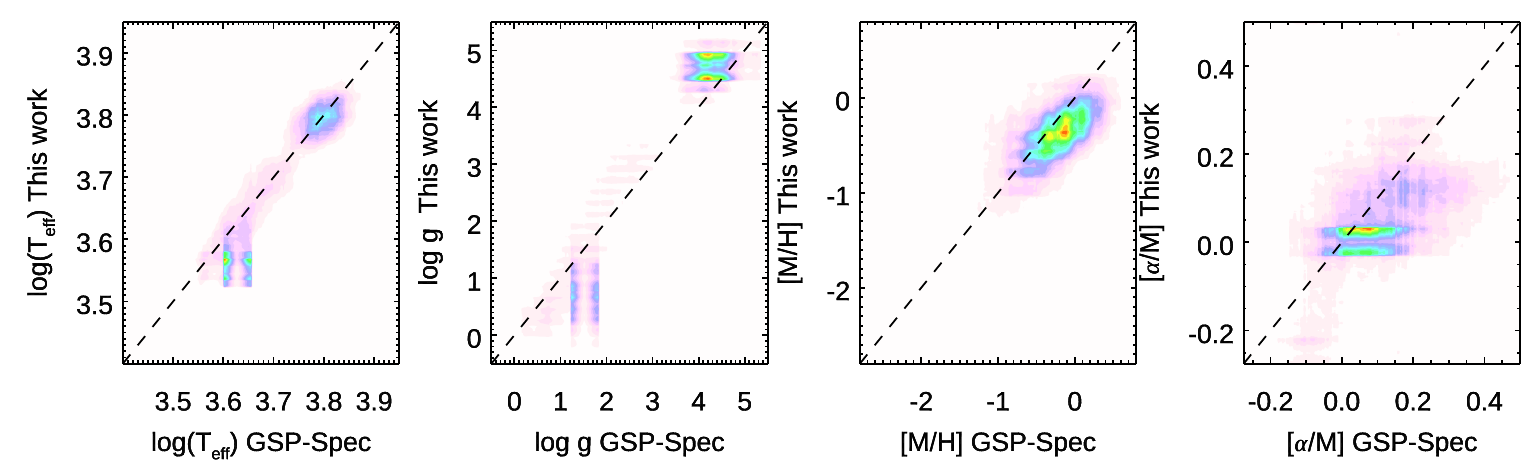}\\
\includegraphics[scale=0.70,angle=0] {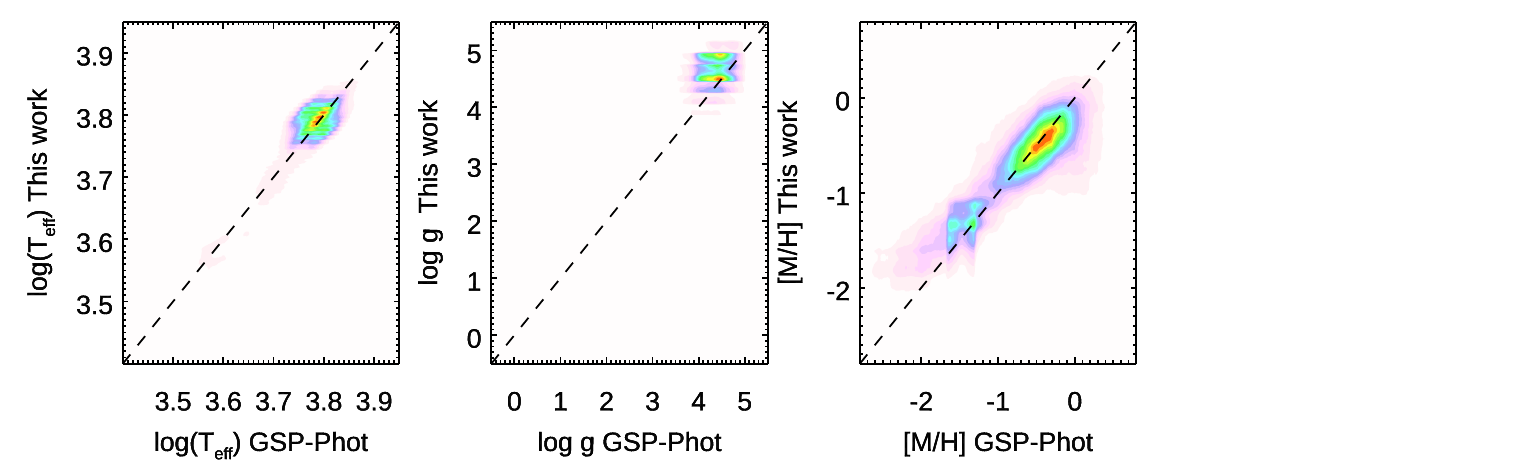}

\caption{Stellar parameter derived from this work compared with the stellar parameters from Gaia DR3.
A density map is generated in each panel to show the distribution in between this work and Gaia DR3 results, where the red color corresponds to the highest density. Top panel: MaStar vs. Gaia DR3 from GSP-Spec. Bottom panel: MaStar vs. Gaia DR3 from GSP-Phot.
The black-dashed lines mark the one-to-one relation to guide the perfect correlation.}
\label{fig:para2gaia} 
\end{figure}

We show the parameter comparison with Gaia's in Fig.~\ref{fig:para2gaia}. 
The mean $\rm T_{eff}$ determined in this work is about 25 K cooler than Gaia GSP-Spec, with an rms of 260 K, and
 is about 43 K warmer than Gaia GSP-Phot, with an rms of 205 K. 
We notice that the over-dense region in the $\rm T_{eff}$ distribution at $\log \rm T_{eff}= 3.8$ is around the temperature range where MaStar has a large number of flux standard stars. The consistency of these stars between Gaia and MaStar ensures the confidence of our stellar parameters.    
We found that there is a vertical stripe in the panel of Gaia GSP-Spec vs. MaStar at $\rm T_{eff} = 4250 K$, and similarly in the panel of surface gravity at $\log g =1.5$.  This may indicate an issue with Gaia GSP-Spec,
possibly due to the pipeline reaching the boundary of the PHOENIX template grid.

Comparable patterns, characterized by vertical stripes in $\rm T_{eff}$ and $\log g$, are also evident in direct comparisons between APOGEE and Gaia GSP-Spec (See Appendix for the detailed density map distribution). This suggests that the observed artifacts are not inherently attributable to the analysis methodology employed in this study. Further discussion is beyond this work. 
The $\log g$, and $\rm [M/H]$ show general consistency between Gaia and MaStar as well, for both  GSP-Spec and GSP-Phot based stellar parameters.  
Similar behavior is observed in the $\rm [M/H]$ between our calculation and GSP-Spec, i.e., our  $\rm [M/H]$  seems to have a systematic shift towards the metal-poor direction compared to GSP-Spec values.  However, we observed excellent consistency in comparing  our $\rm [M/H]$  to GSP-Phot, especially at the over-dense area (see the lower panel of Fig.~\ref{fig:para2gaia}). We provide the metallicity conversions between MaStar and Gaia below. Users can decide which to use for metallicity: $\rm [M/H]$ or the converted metallicities, i.e., $\rm  [M/H]_{ToAPO}$, $\rm  [M/H]_{ToGspec}$ and $\rm  [M/H]_{ToGphot}$.


\begin{equation} \label{eq:2}
{\rm [M/H]_{ToGspec} =-0.1397 \times {\rm [M/H]_{MaStar}} ^2 +0.8214 \times {\rm [M/H]_{MaStar}} + 0.1358   }   
\end{equation}
  
\begin{equation}  \label{eq:3}
{\rm [M/H]_{ToGphot} =-0.1951 \times {\rm [M/H]_{MaStar}} ^2 +0.7927 \times {\rm [M/H]_{MaStar}} - 0.0336 }
\end{equation}

The median uncertainty of our {$\rm [M/H]$} measurements is 0.32 dex. The median uncertainty of {$\rm [M/H]$} from GSP-Spec is 0.21 dex. The median uncertainty of {$\rm [M/H]$} from GSP-Phot  is 0.06 dex. The RMS of {$\rm [M/H]$} residual relative to the above fitting equation \eqref{eq:2} for  GSP-Spec is 0.35 dex. 
The RMS of {$\rm [M/H]$} residual relative to the above fitting equation \eqref{eq:3} for GSP-Phot is 0.37 dex. Therefore, the uncertainties of  {$\rm [M/H]$} from  GSP-Phot is likely to be underestimated, as the RMS of residuals between our {$\rm [M/H]$} and GSP-Phot is larger than what the individual median uncertainties could contribute to the total RMS, unlike the {$\rm [M/H]$} of our calculation and GSP-Spec that are consistent within their uncertainties.


For the convenience of the users, we also offer the reverse formulae from APOGEE, GSP-Spec and GSP-Phot to MaStar parameters:
 \begin{equation}
{\rm {[M/H]}_{ToMaS} =0.0860  \times {\rm [M/H]_{APO}} ^2   +0.9029 \times {\rm [M/H]_{APO}}   -0.2022.}   
 \end{equation}

 \begin{equation}
{\rm {[M/H]}_{ToMaS}  =-0.0817 \times {\rm {[M/H]}_{Gspec}} ^2   +0.6538 \times {\rm {[M/H]}_{Gspec}} -0.2483.}
 \end{equation}
 
 \begin{equation}
{\rm {[M/H]}_{ToMaS}  =0.0309 \times {\rm {[M/H]}_{Gphot}} ^2   +0.8641 \times {\rm {[M/H]}_{Gphot}} -0.1018.}
 \end{equation}

The $\rm [\alpha/M]$ parameter between Gaia GSP-Spec and this work are 
consistent. The RMS scatter in the Y-direction is 0.21 dex, which is consistent with the combination of the median uncertainty of our measurements (0.18 dex) and that of Gaia GSP-Spec (0.12 dex).
We noticed there are some similar artifacts of the  $\rm [M/H]$  and  $\rm [\alpha/M]$ in Fig.~\ref{fig:para2apo} and Fig.~\ref{fig:para2gaia}.
For instance, the horizontal features in the right panels at $\rm [\alpha/M]$ = 0 could be due to the fact that our solar-$\rm [\alpha/M]$ models are denser than the non-solar-$\rm [\alpha/M]$ models, leading to a concentration of points around $\rm [\alpha/M]\sim 0$. 
 In order to test if the artifacts of the  $\rm [M/H]$  and  $\rm [\alpha/M]$ plot originate from this paper's analysis, 
we directly compared APOGEE and Gaia's results. These are shown in Fig.~\ref{fig:apo2gaia}  and described in Appendix~\ref{appendix:apogaia}.  
The systematic difference between our $\rm [M/H]$ and theirs could be due to the use of different templates. 
What we found was that $\rm [M/H]$ in between GSP-Spec and APOGEE suggests consistency. 
We noted that both APOGEE and Gaia are mainly using the template from MARCS synthetic atmospheric models, while our parameters are dominated from the template of BOSZ. That could be the potential reason of metallicity distribution deficits.

\section{Conclusions}\label{secconclusion}

We present a new method that complements our team's previous efforts to determine stellar parameters for the MaStar stellar library.
We assign priors using theoretical isochrones, which contain Gaia magnitude information comparable to that of MaStar, where a fitting-subgrid is defined to reduce computation time. Full-spectrum fitting between the data and model spectra is performed to evaluate the fitting-quality of each candidate-model. The continuum is modeled with a multiplicative polynomial. 
Due to the large stellar parameter coverage of MaStar, we use theoretical atmospheric models (BOSZ and MARCS) as spectral templates. Specifically, we use the fine version of BOSZ grids through private collaboration. Therefore, no interpolation of the model spectra between grid points is performed in the full-spectrum fitting.
Our stellar parameters are estimated using a Bayesian average, namely, likelihood-weighted stellar parameters  
with a flat prior derived from the Gaia color-magnitude diagram within the subgrid.


We select a few BOSZ models and their respective stellar parameters as mock spectra to test the recovery of parameters using this method. We find that these agree within one-sigma.
Our stellar parameters are consistent with several literature catalogs, including APOGEE and Gaia DR3. The comparison among parameters within the MaStar working group will be shown in \citet{Yan24}. Moreover, we perform internal stellar parameter consistency checks using repeated observations. These internal consistency checks show excellent agreement for each of 
the four stellar parameters -  $\rm T_{eff}$, $\log g$, [M/H], and $\rm [\alpha/M]$. 


There are some ``quantized" $\log g$ values which mostly occurs in stars with temperature $\rm T_{eff} \leq 9000K$. The source of that feature may come from averaging two candidate $\log g$ values next to each other satisfying the fitting procedure. The artifact in the right panel of Fig.~\ref{fig:paramscomb} at $\rm [\alpha/M]=0$ could be improved by involving finer non-solar models, as the steps of our non-solar models are larger than the solar models.


Since several factors
may contribute to the final stellar parameters, we assign parameter quality flags to help identify the most reliable values. We present a sample of stellar parameters in Table~\ref{sampletable}, where `flag\_1 = 1' marks  reliable values for the three main stellar parameters $\rm T_{eff}$, $\log g$ and [M/H],  and `flag\_2 = 1' marks a reliable value for the $\rm [\alpha/M]$ estimate. In addition, we indicate the source of the best-fit template, either from BOSZ or from MARCS. Our results are part of the MaStar team work. The parameter catalog is 
available in the SDSS-IV DR17 data release as a value-added catalog{\footnote{\url{https://www.sdss4.org/dr17/mastar/mastar-stellar-parameters/}}}. 

\appendix

\section{Stellar parameters of APOGEE and Gaia DR3}\label{appendix:apogaia}

\begin{figure}
\includegraphics[scale=0.52,angle=0] {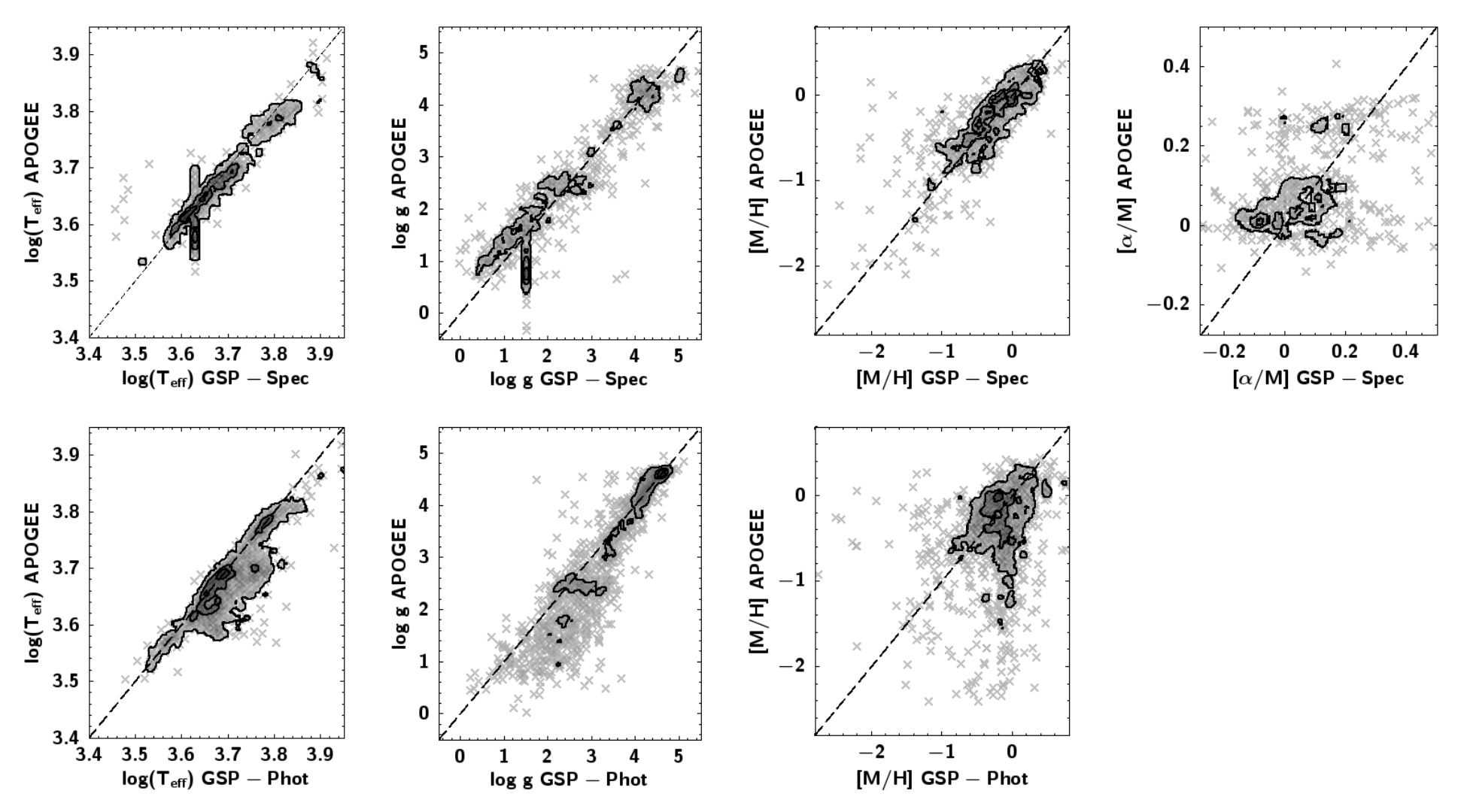}
\caption{Stellar parameter derived from APOGEE compared with the stellar parameters from Gaia DR3.
A density map is generated in each panel to show the distribution in between APOGEE and Gaia DR3 results. Top panel: APOGEE vs. Gaia DR3 from GSP-Spec. Bottom panel: APOGEE vs. Gaia DR3 from GSP-Phot.
The black-dashed lines mark the one-to-one relations to guide the perfect correlation.}
\label{fig:apo2gaia} 
\end{figure}

We investigated the consistency of stellar parameters derived in the literature from APOGEE and Gaia DR3. 
The direct parameter comparison in Fig.~\ref{fig:apo2gaia} shows that the vertical artifacts in $\rm T_{eff}$ and $\rm \log g$ in GSP-Spec exhibit the same features as those in Fig.~\ref{fig:para2gaia}. Therefore, these vertical artifacts are unlikely to originate from the analysis presented in this work. The [M/H] distribution shows good agreement between APOGEE and Gaia DR3 from GSP-Spec, which could be attributed to the fact that they use the same theoretical templates. In contrast, the parameter comparison of [M/H] between APOGEE and Gaia GSP-Phot shows a larger scatter, as expected. 

The RMS differences between APOGEE and GSP-Phot are 0.04 for $\log T_{\rm eff}$, 0.50 for $\log g$, 0.48 for [M/H], and 0.16 for $\rm [\alpha/M]$. In comparison, the RMS differences between APOGEE and GSP-Spec are 0.05 for $\log T_{\rm eff}$, 0.56 for $\log g$, and 0.59 for [M/H].

\section{Potential stellar parameter degeneracy}\label{appendix:paramdis}

\begin{figure}
\centering
\includegraphics[scale=0.4,angle=0]{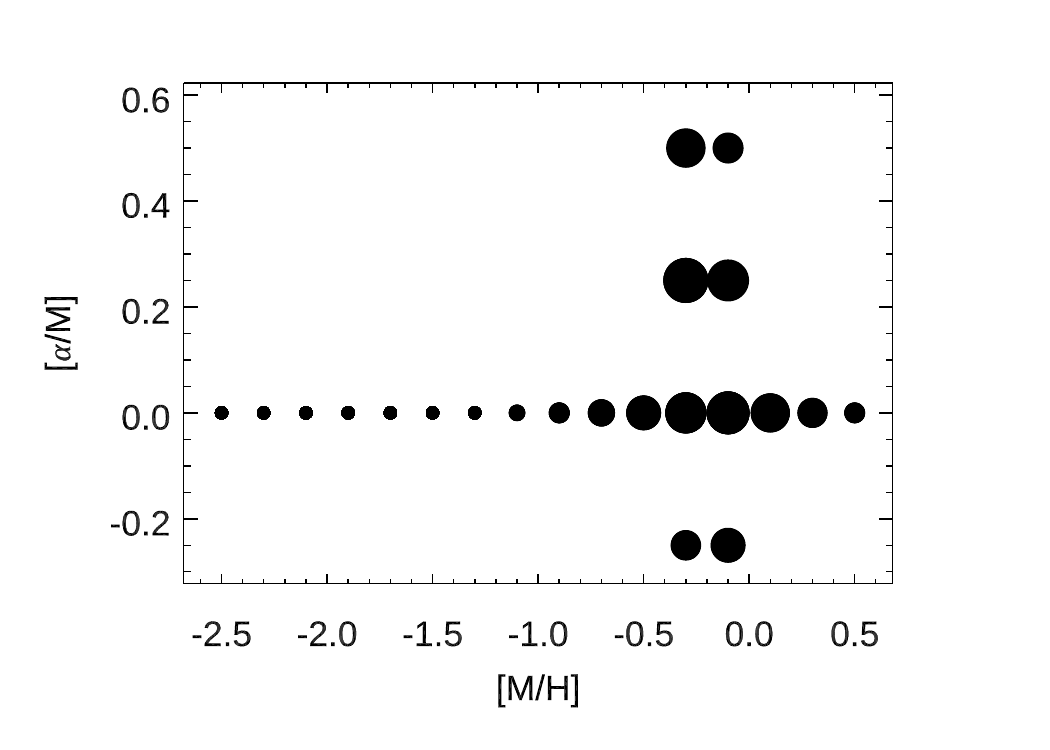}
\includegraphics[scale=0.4,angle=0]{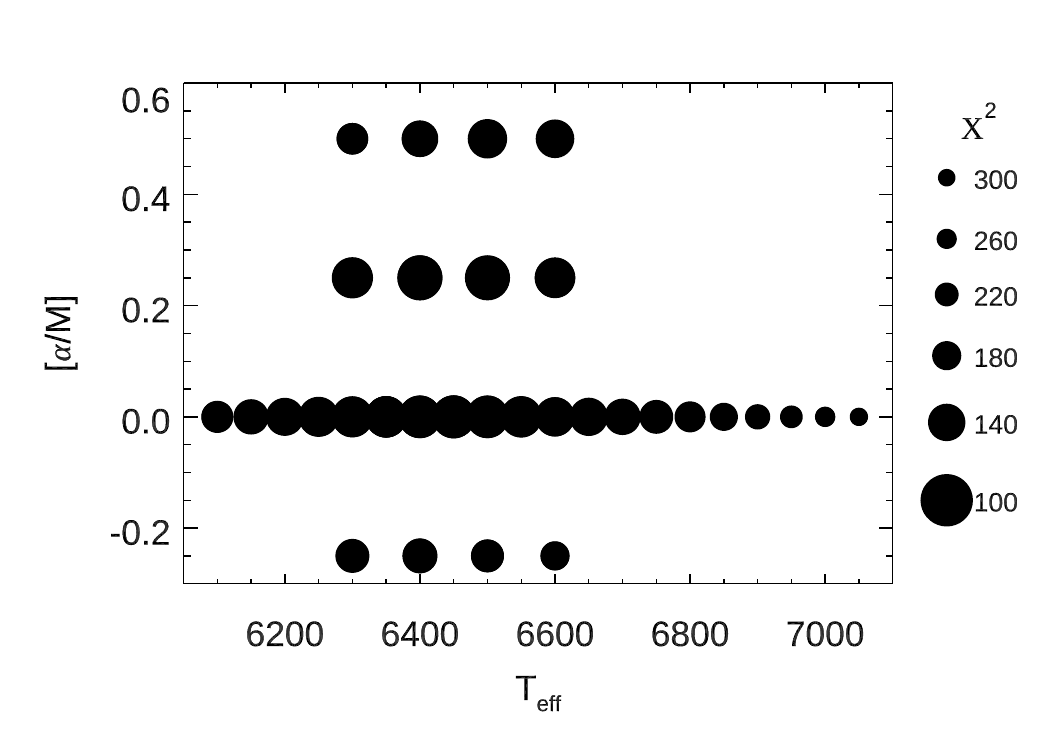}\\
\includegraphics[scale=0.4,angle=0]{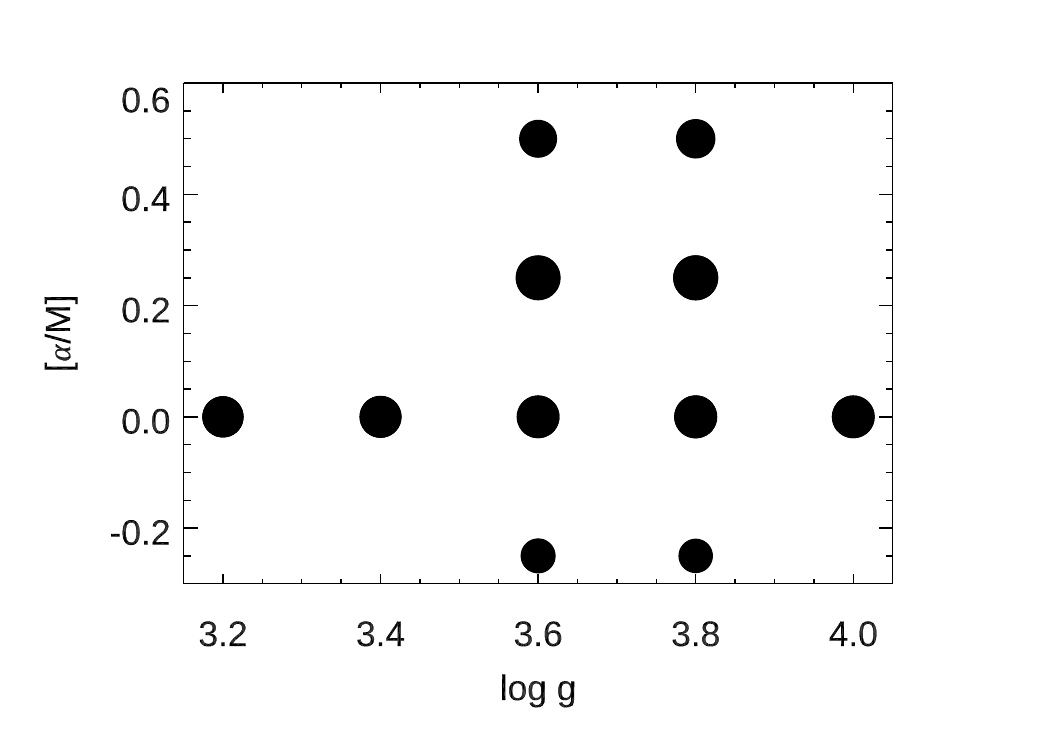}
\includegraphics[scale=0.4,angle=0]{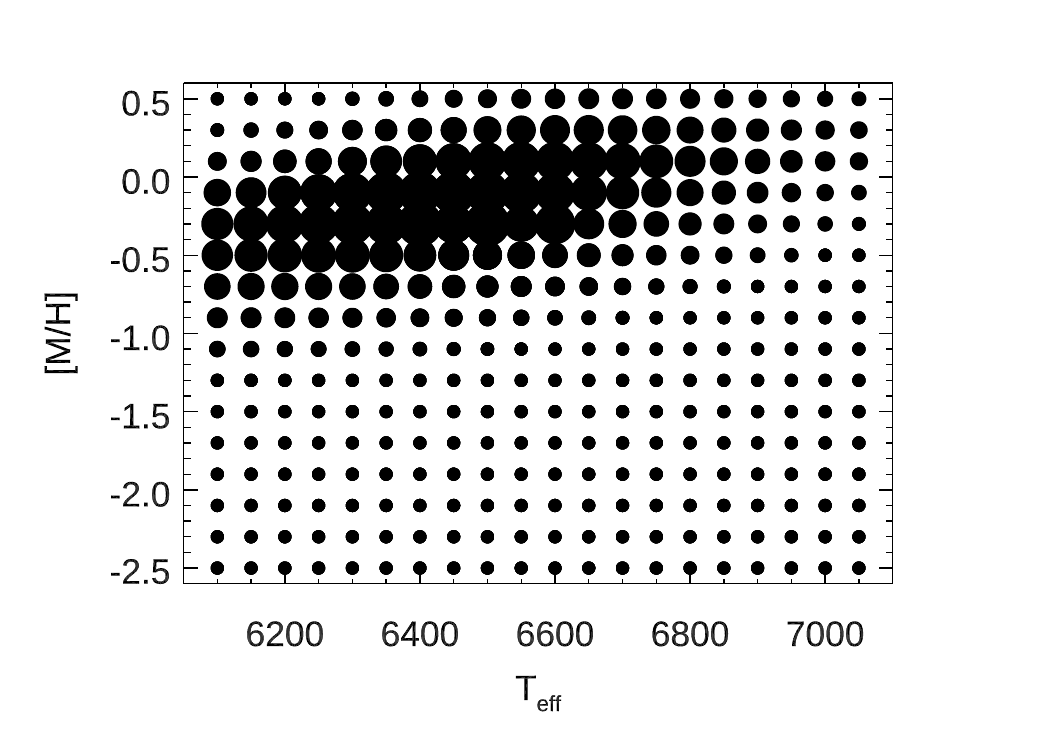}\\
\advance\leftskip0.3cm\includegraphics[scale=0.4,angle=0]{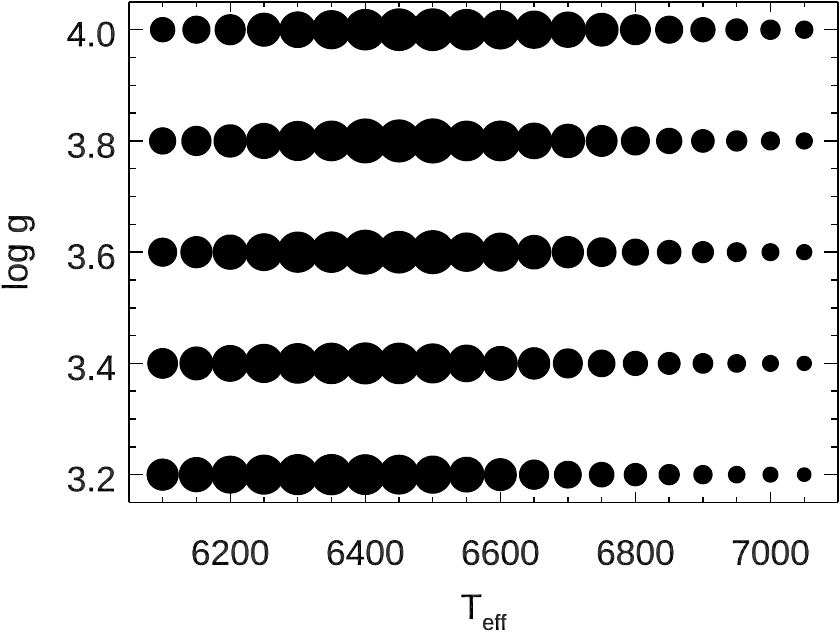}
\hspace*{1cm}\includegraphics[scale=0.4,angle=0]{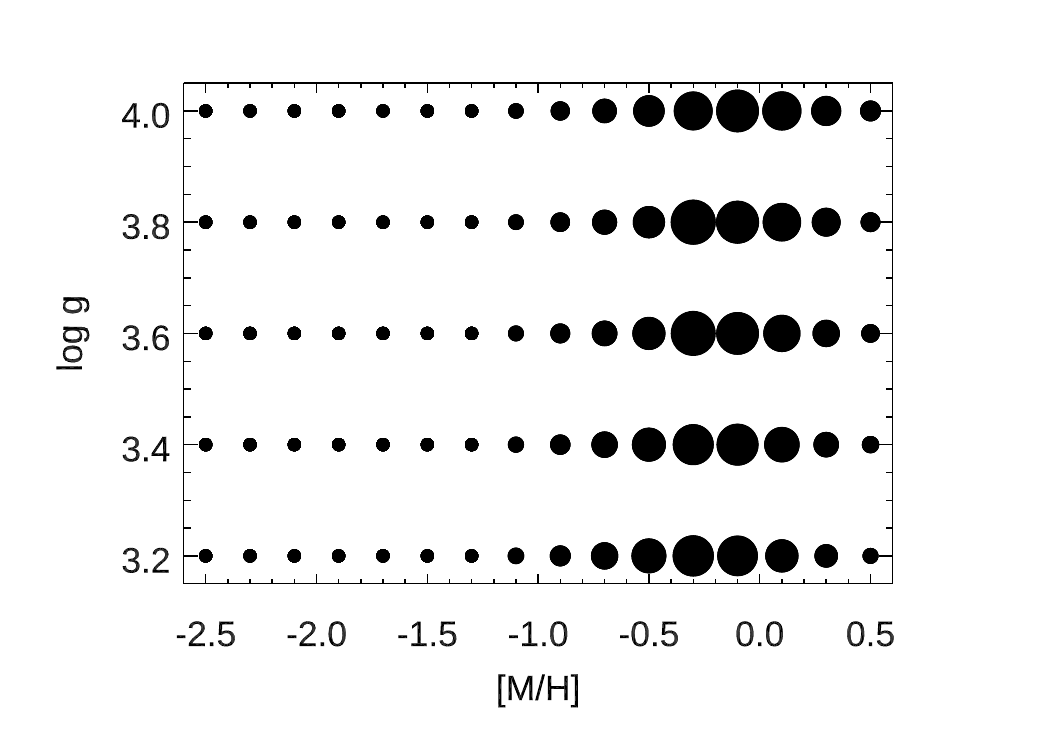}
\caption{ Stellar parameters candidates for MANGAID=60-342923456068249472: $\rm T_{eff}=6439.1K, logg=3.73$, $\rm [M/H]=-0.3$, $[\alpha/M]=0.25$. Symbol sizes are scaled in proportion to $1/\chi^2$ with the scale shown in the legend to the right of the upper right panel. } 
\label{fig:paramcomp1}
\end{figure}

\begin{figure}
\centering
\includegraphics[scale=0.4,angle=0]{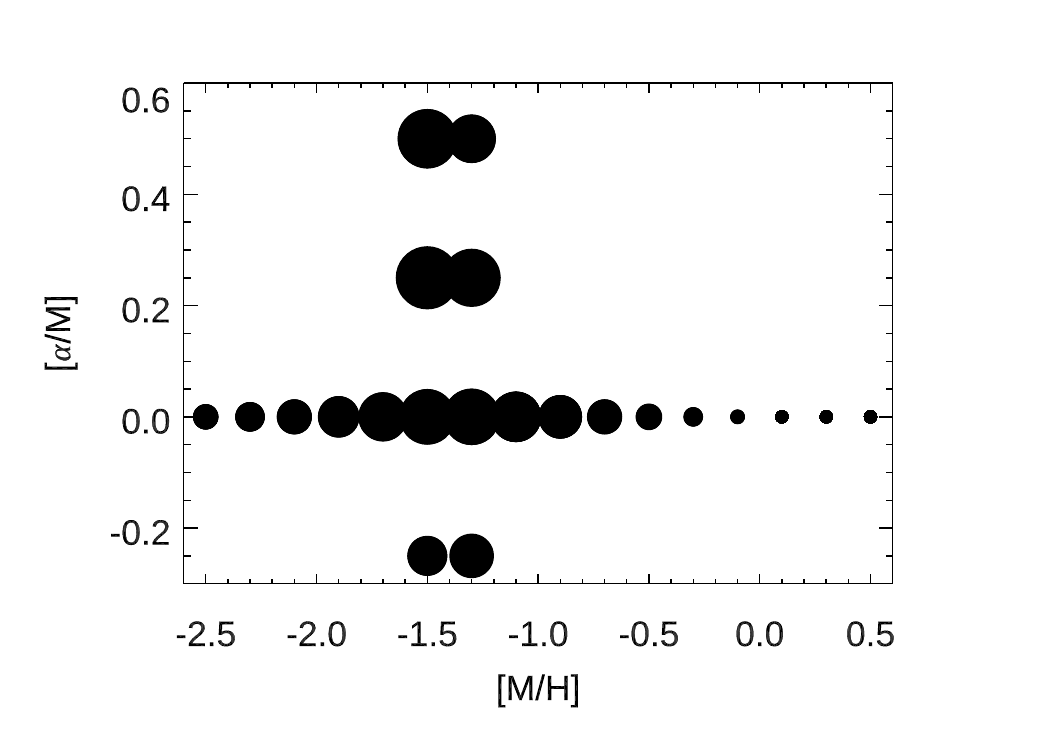}
\includegraphics[scale=0.4,angle=0]{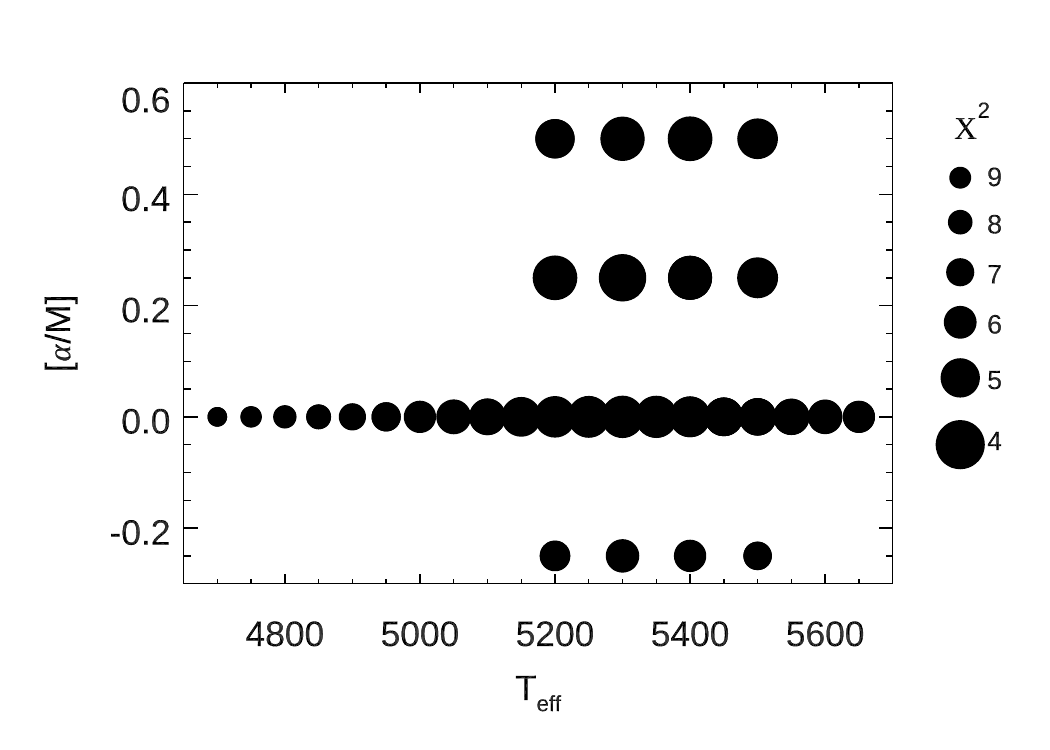}\\
\includegraphics[scale=0.4,angle=0]{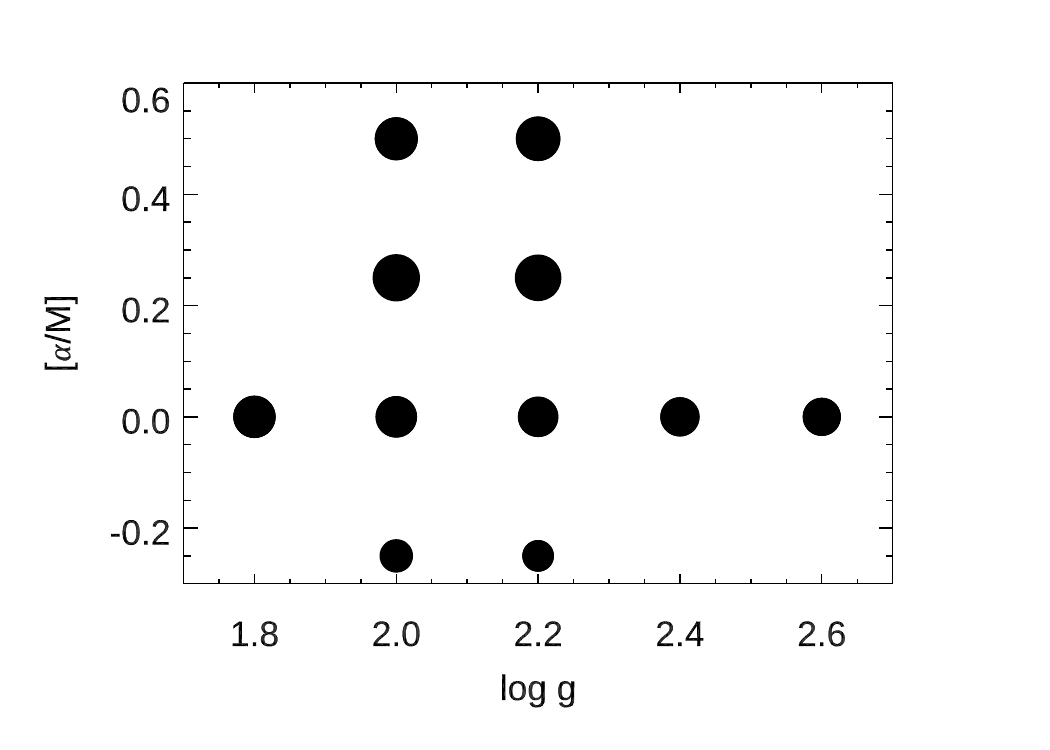}
\includegraphics[scale=0.4,angle=0]{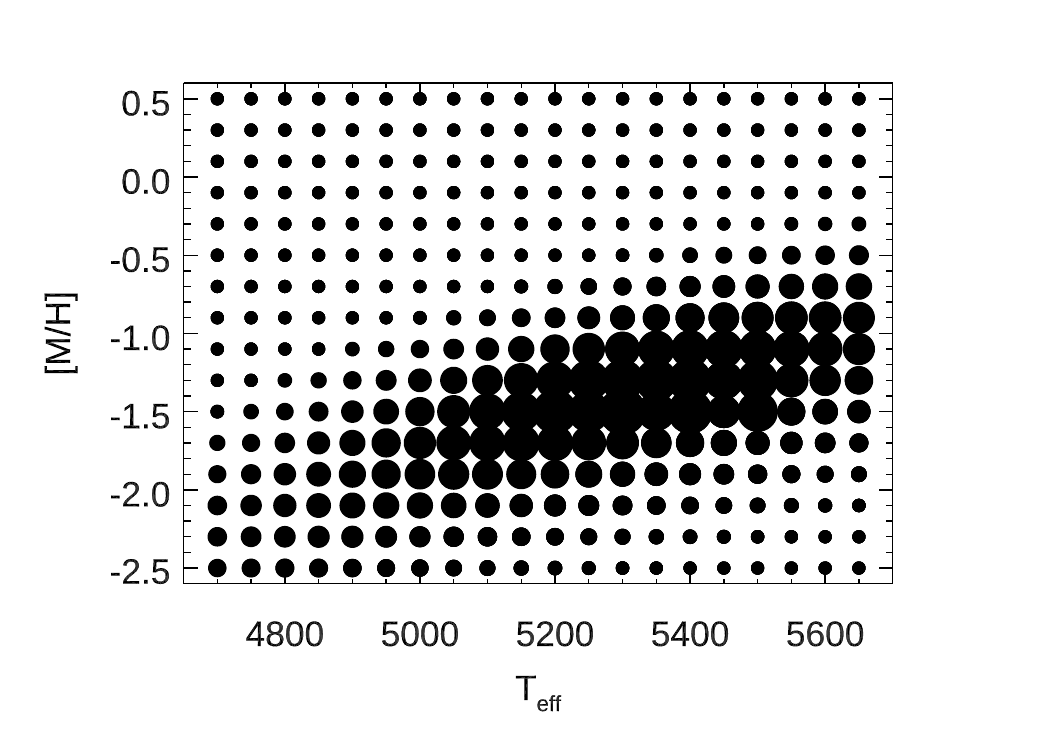}\\
\includegraphics[scale=0.4,angle=0]{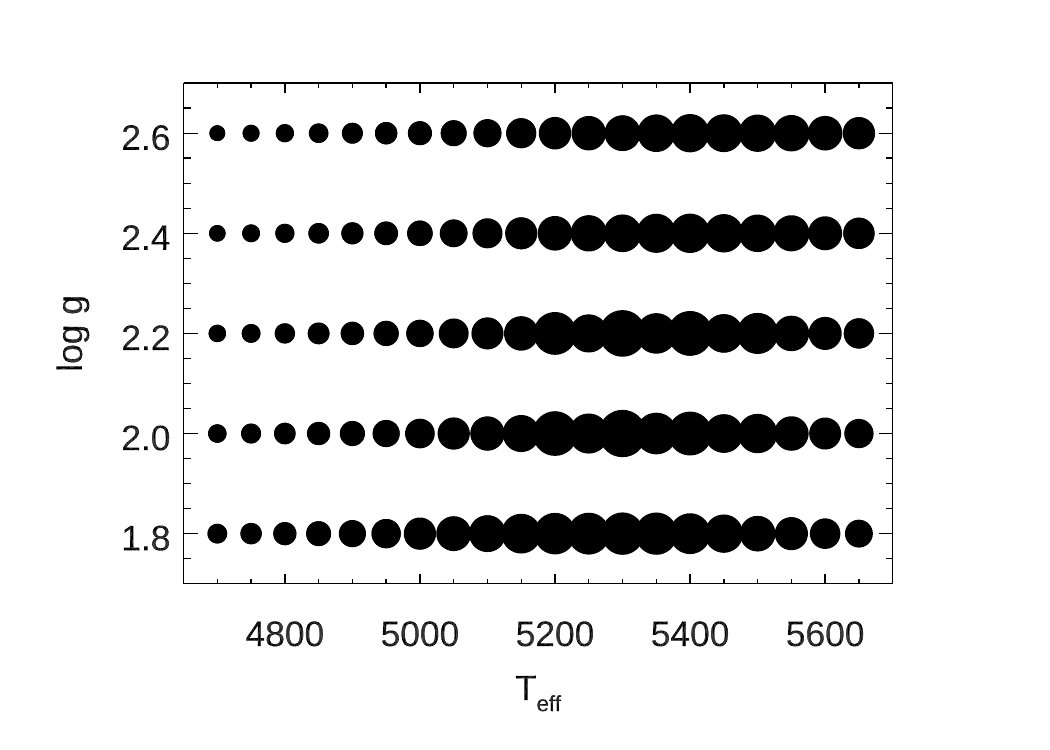}
\includegraphics[scale=0.4,angle=0]{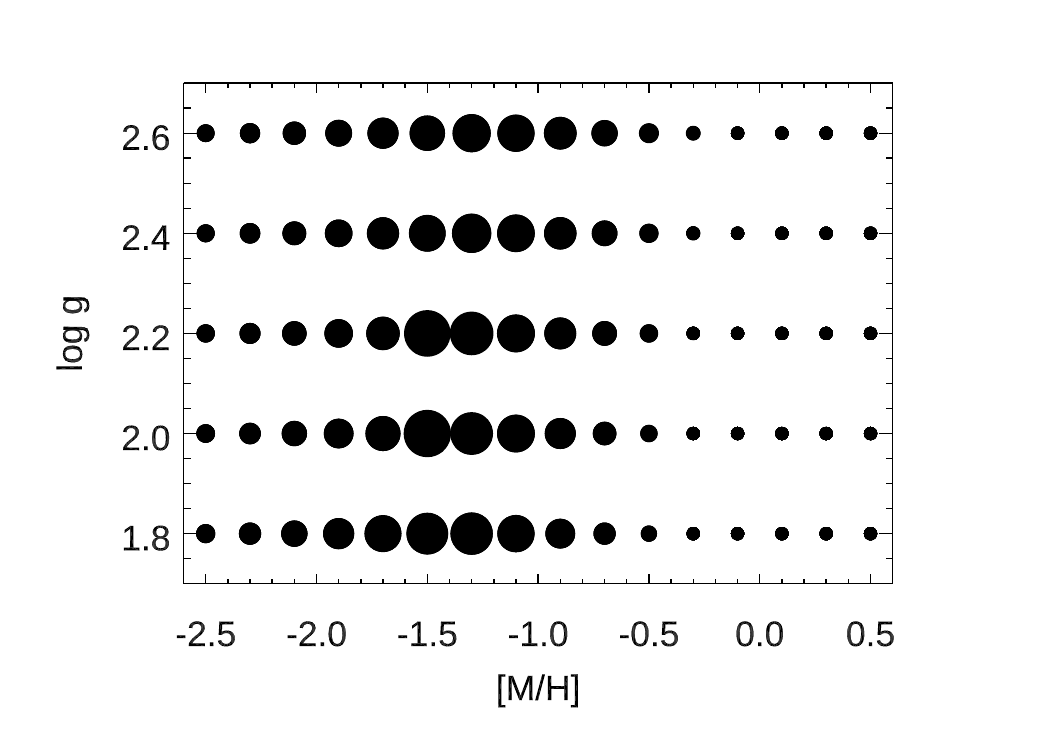}
\caption{Stellar parameters candidates for  $\rm MANGAID=60-3608087611736692992\ T_{eff}=5317.636K, logg=1.90, [M/H]=-1.54, [\alpha/M]=0.27$. Symbol sizes are scaled in proportion to $1/\chi^2$ with the scale shown in the legend to the right of the upper right panel. }
\label{fig:paramcomp2}
\end{figure}

For each individual star, we computed the goodness of fit of each candidate model against its stellar parameters. Subsequently, Bayesian averaging was employed to derive the final parameter sets. To facilitate readers' comprehension of the procedure, we present illustrative examples of the Chi-square distribution for two stars in Fig.~\ref{fig:paramcomp1} and Fig.~\ref{fig:paramcomp2}: MANGAID=60-342923456068249472 $\rm T_{eff}=6439.1K, logg=3.73$, $\rm [M/H]=-0.3$, $[\alpha/M]=0.25$, and  MANGAID=60-3608087611736692992, $\rm T_{eff}=5317.636K, logg=1.90, [M/H]=-1.54, [\alpha/M]=0.27$, respectively. The size of the symbols is proportional to the goodness of the fit, i.e., $1/\chi^2$. As seen from the example plots, we observe no strong degeneracy between $\rm  [M/H]$ and $\rm [\alpha/M]$.

\begin{sidewaystable}
\caption{Stellar Parameters Adopted in this Work} 
\hspace*{-3.5\baselineskip} 
 \begin{threeparttable}\label{sampletable}
 \footnotesize 
\begin{tabular}{|l|l|l|l|l|l|r|r|l|l|l|l|l|l|}
\hline
  \multicolumn{1}{c|}{OBNAME} &
  \multicolumn{1}{|c|}{MANGAID} &
  \multicolumn{1}{c|}{$\rm T_{eff}$} &
  \multicolumn{1}{c|}{$\log g$} &
  \multicolumn{1}{c|}{$\rm [M/H]$} &
  \multicolumn{1}{c|}{$\rm [\alpha/M]$} &
  \multicolumn{1}{c|}{flag\_1} &
  \multicolumn{1}{c|}{flag\_2} &
  \multicolumn{1}{c|}{SOURCE} &
  \multicolumn{1}{c|}{$\rm T_{eff}$ err} &
  \multicolumn{1}{c|}{$\log g$ err} &
  \multicolumn{1}{c|}{$\rm [M/H]$ err} &
  \multicolumn{1}{c|}{$\rm [\alpha/M]$ err} &
  \multicolumn{1}{c|}{$\chi^2$} \\
 &  & (K) &   & dex & dex &\  &\   &\  & (K) &   & dex & dex &  \\
\hline
   9475-6104-57732 & 3-51508760 &5452. & 4.90 & -0.69 & +0.16 & 1 & 1 & bosz & 110 & 0.10 & 0.33 & 0.27 & 1.23\\
   9475-6104-57734 & 3-51508760 & 5536. & 4.70 & -0.59 & +0.16 & 1 & 0 & bosz & 111 & 0.10 & 0.34 & 0.28 & 1.38\\
  9477-12705-57796 & 3-51508785 & 5752. & 4.71 & -0.21 & +0.05 & 1 & 1 & bosz & 107 & 0.10 & 0.25 & 0.24 & 3.46\\
  9476-1902-57764 & 3-51509189 & 3825. & 4.32 & +0.46 & +0.07 & 1 & 1 & marcs & 99 & 0.28 & 0.35 & 0.18 & 108.02\\
  9469-12704-58033 & 3-51509589 & 6158. & 4.70 & -0.64 & +0.14 & 1 & 1 & bosz & 104 & 0.10 & 0.27 & 0.25 & 3.80\\
  9469-12704-58037 & 3-51509589 & 6161. & 4.70 & -0.63 & +0.13 & 1 & 1 & bosz & 105 & 0.10 & 0.28 & 0.25 & 3.21\\
  8894-3702-57433 &3-51509632 &  3990. & 4.74 & +0.81 & +0.12 & 1 & 1 & marcs & 106 & 0.29 & 0.36 & 0.22 & 51.80\\
 9468-3701-57794 & 3-51509774 & 3967. & 4.48 & -1.45 & +0.34 & 1 & 1 & marcs & 247 & 0.40 & 0.76 & 0.38 & 2.90\\
  9468-3701-57795 & 3-51509774 & 3937. & 4.48 & -1.34 & +0.30 & 1 & 1 & marcs & 208 & 0.40 & 0.67 & 0.34 & 3.68\\
  8894-703-57433 & 3-51509897 & 5394. & 4.47 & -1.64 & +0.39 & 1 & 0 & bosz & 170 & 0.40 & 0.47 & 0.40 & 4.60\\
\hline\end{tabular}

   \begin{tablenotes}

      \small
     \hspace*{+12.5\baselineskip} 
  \begin{minipage}{18.5cm}

      \item A sample of the stellar parameters from this work. The 1-sigma uncertainty is taken as each parameter's error. The reduced-$\chi^2$ is obtained from the full-spectrum-fitting of the best weighted theoretical atmospheric model and the MaStar spectra. The column ``OBNAME" is a combination of ``PLATE-IFUDESIGN-MJD". The column `flag\_1 = 1' indicates reliable measurements of the three primary stellar parameters: $\rm T_{eff}$, $\log g$, and [M/H]. Similarly, `flag\_2 = 1' indicates a reliable estimate of $\rm [\alpha/M]$. The full table is available in digital format only. 
  \end{minipage}%
    \end{tablenotes}
  \end{threeparttable}

\end{sidewaystable}


\acknowledgments

\section*{acknowledgments}

Y. C. acknowledges the support of NYU Abu Dhabi AD013.  
R. Y. acknowledges support by grants from the National Natural Science Foundation of China (NSFC; grant No. 12425302), by the Hong Kong Jockey Club Charities Trust through the JC STEM Lab of Astronomical Instrumentation, and by grants from the Research Grants Council of the Hong Kong Special Administrative Region, China [Project No: CUHK 14302522, 14303123], and by the Direct Grant of CUHK Faculty of Science.  The research of J. D. G. is supported by NYU Abu Dhabi Grant AD022.

This research was carried out on the High Performance Computing resources at New York University Abu Dhabi. This material is based upon work supported by Tamkeen under the NYU Abu Dhabi Research Institute grant CASS.

This project made use of data taken in SDSS-IV. 

Funding for the Sloan Digital Sky Survey IV has been provided by the Alfred P. Sloan Foundation, the U.S. Department of Energy Office of Science, and the Participating Institutions. SDSS-IV acknowledges
support and resources from the Center for High-Performance Computing at the University of Utah. The SDSS web site is {\url{www.sdss.org}}.

SDSS-IV is managed by the Astrophysical Research Consortium for the 
Participating Institutions of the SDSS Collaboration including the 
Brazilian Participation Group, the Carnegie Institution for Science, 
Carnegie Mellon University, the Chilean Participation Group, the French Participation Group, Harvard-Smithsonian Center for Astrophysics, 
Instituto de Astrof\'isica de Canarias, The Johns Hopkins University, Kavli Institute for the Physics and Mathematics of the Universe (IPMU) / 
University of Tokyo, the Korean Participation Group, Lawrence Berkeley National Laboratory, 
Leibniz Institut f\"ur Astrophysik Potsdam (AIP),  
Max-Planck-Institut f\"ur Astronomie (MPIA Heidelberg), 
Max-Planck-Institut f\"ur Astrophysik (MPA Garching), 
Max-Planck-Institut f\"ur Extraterrestrische Physik (MPE), 
National Astronomical Observatories of China, New Mexico State University, 
New York University, University of Notre Dame, 
Observat\'ario Nacional / MCTI, The Ohio State University, 
Pennsylvania State University, Shanghai Astronomical Observatory, 
United Kingdom Participation Group,
Universidad Nacional Aut\'onoma de M\'exico, University of Arizona, 
University of Colorado Boulder, University of Oxford, University of Portsmouth, 
University of Utah, University of Virginia, University of Washington, University of Wisconsin, 
Vanderbilt University, and Yale University.


\begin{thebibliography}{}

\bibitem[Abdurro'uf et al.(2022)]{Abdurro'uf22}Abdurro’uf, A. K., Aerts, C., et al. 2022, ApJS, 259, 35
\bibitem[Allende Prieto et al.(2018)]{Allende18}Allende Prieto, C., Koesterke, L., Hubeny, I., et al. 2018, A\&A, 618, A25
\bibitem[Anders \& Grevesse(1989)]{anders89} Anders, E., \& Grevesse, N. 1989, Geochim. Cosmochim. Acta, 53, 197
\bibitem[Asplund et al.(2005)]{asplund05} Asplund, M., Grevesse, N., \& Sauval, A. J. 2005, in Astronomical Society of the Pacific 
Conference Series, Vol. 336, Cosmic Abundances as Records of Stellar Evolution and Nucleosynthesis, ed. T. G. Barnes, III \& F. N. Bash, 25
\bibitem[Asplund et al.(2009)]{asplund09} Asplund, M., Grevesse, N., Sauval, A. J., \& Scott, P. 2009, ARA\&A, 47, 481
\bibitem[Beers et al.(1990)]{Beers1990}{Beers}, Timothy C., {Flynn}, Kevin, {Gebhardt}, Karl, 1990, AJ, 100, 32B
\bibitem[Bohlin et al.(2017)]{bohlin17} Bohlin, R. C., M{\'e}sz{\'a}ros, S., Fleming, S. W., Gordon, K. D. Koekemoer, A. M., 
Kov{\'a}cs, J. 2017, \aj, 153, 234
\bibitem[Bressan et al.(2012)]{Bressan12}{Bressan}, Alessandro,  {Marigo}, Paola,  {Girardi}, L{\'e}o. et al., 2012, MNRAS, 427, 127B
\bibitem[Bohlin et al.(2017)]{BOSZ}Bohlin, R. C., {M{\'e}sz{\'a}ros}, S., Fleming, S. ~W., et al. 2017, AJ, 153, 234B
\bibitem[{Bruzual} \& {Charlot}(2003)]{BC03} {Bruzual}, G. \& {Charlot}, S. 2003, MNRAS, 344, 1000B
\bibitem[Bundy et al.(2015)]{Bundy15}{Bundy}, Kevin, {Bershady}, Matthew A., {Law}, David R., et al., 2015, ApJ, 798, 7B
\bibitem[Cenarro et al.(2001)]{Cenarro01}Cenarro, A. J., Cardiel, N., Gorgas, J., et al., 2001, MNRAS, 326, 959,
\bibitem[Chen Y-P et al.(2014)]{XSL}Chen, Yan-Ping, Trager, S. C., Peletier, R. F., et al. 2014, A\&A, 565A,117C
\bibitem[Chen et al.(2014)]{Chen14}{Chen}, Yang,  {Girardi}, L{\'e}o,  {Bressan}, Alessandro et al., 2014, MNRAS , 444, 2525C 
\bibitem[Chen et al.(2015)]{Chen15}{Chen}, Yang,  {Bressan}, Alessandro, {Girardi}, L{\'e}o et al., 2015, MNRAS, 452, 1068C
\bibitem[Chen et al.(2020)]{Chenparam20}Chen Y.-P., Yan R.,  Maraston C., et al., 2020, ApJ, 899, 62
\bibitem[Cirasuolo et al.(2014)]{Cirasuolo14}Cirasuolo, M.,  Afonso, J.,  Carollo, M., et al., 2014, SPIE, 9147E, 0NC
\bibitem[Coelho et al.(2005)]{Coelho05}{Coelho}, P., {Barbuy}, B., {Mel{\'e}ndez}, J., {Schiavon}, R.~P., {Castilho}, B.~V., 2005, A\&A, 443, 735C
\bibitem[Coelho et al.(2007)]{Coelho07}Coelho, P., Bruzual, G., Charlot, S., et al., 2007, MNRAS, 382, 498
\bibitem[Coelho(2014)]{coelho14} Coelho, P. R. T. 2014, \mnras, 440, 1027
\bibitem[Conroy(2013)]{Conroy13}Conroy, Charlie, 2013, ARA\&A, 51, 393
\bibitem[Creevey et al.(2022)]{Creeveygaia22}{Creevey}, O.~L. and {Sordo}, R. and {Pailler}, F. et al., 2022, arXiv: 2206.05864C
\bibitem[Cui et al.(2012)]{cui12}Cui, X.-Q., Zhao, Y.-H., Chu, Y.-Q., et al. 2012, Research in Astronomy and Astrophysics, 12, 1197, doi: 10.1088/1674-4527/12/9/003
\bibitem[de Jong et al(2019)]{dejong19}de Jong R. S., et al., 2019, Msngr, 175, 3
\bibitem[de Laverny et al.(2012)]{deLaverny12}de Laverny, P., Recio-Blanco, A., Worley, C. C.,, Plez, B. 2012, A\&A, 544, A126
\bibitem[De Silva et al.(2015)]{deSilva15}De Silva, G. M., Freeman, K. C., Bland-Hawthorn, J., et al. 2015, MNRAS, 449, 2604
\bibitem[Deng et al.(2012)]{deng12}Deng, L.-C., Newberg, H. J., Liu, C., et al. 2012, Research in Astronomy and Astrophysics, 12, 735, doi: 10.1088/1674-4527/12/7/003
\bibitem[Diaz et al.(1989)]{Diaz89}{Diaz}, A.~I.,  {Terlevich}, E., {Terlevich}, R., 1989, MNRAS, 239, 325D
\bibitem[Drory et al.(2015)]{Drory15}{Drory}, N.,  {MacDonald}, N., {Bershady}, M.~A. et al., 2015, AJ, 149, 77
\bibitem[Driver et al.(2019)]{Driver19}{Driver}, S.~P.,  {Liske}, J., {Davies}, L.~J.~M. et al. 2019, Msngr, 175, 46D
\bibitem[Falc{\'o}n-Barroso et al.(2011)]{Falcon-barroso11}{Falc{\'o}n-Barroso}, J., {S{\'a}nchez-Bl{\'a}zquez}, P., {Vazdekis}, A., et al., 2011, A\&A, 532A, 95F
\bibitem[Fouesneau et al.(2022)]{Fouesneaugaia22}{Fouesneau}, M. and {Fr{\'e}mat}, Y. and {Andrae}, R. et al., 2022, arXiv: 2206.05992F
\bibitem[Gaia Collaboration et al.(2021)]{gaiadr3}Gaia Collaboration et al., 2021, A\&A, 649A, 1G
\bibitem[Garc{\'\i}a P{\'e}rez et al.(2016)]{Gacia16} Garc{\'\i}a P{\'e}rez, Ana E.,  {Allende Prieto}, Carlos,  {Holtzman}, et al., 2016, AJ, 151, 144G
\bibitem[Gilmore et al.(2022)]{Gilmore22}{Gilmore}, G., {Randich}, S., {Worley}, C.~C. et al., 2022A\&A, 666A, 120G
\bibitem[Gregg et al.(2006)]{ngsl}Gregg, M. D., Silva, D., Rayner, J., et al. 2006, hstc, conf, 209G
\bibitem[Grevesse \& Sauval(1998)]{grevesse98} Grevesse, N., \& Sauval, A. J. 1998, Space Sci. Rev., 85, 161
\bibitem[Grevesse et al.(2007)]{grevesse07} Grevesse, N., Asplund, M., \& Sauval, A. J., 2007, Space Sci. Rev., 130, 105
\bibitem[Grevesse \& Sauval(1998)]{Grevesse98}Grevesse, N., \& Sauval, A. J. 1998, Space Sci. Rev., 85, 161
\bibitem[Gunn et al.(2006)]{Gunn06}Gunn J. E., et al., 2006, AJ, 131, 2332
\bibitem[Gustafsson et al.(2008)]{Gustafsson08}{Gustafsson}, B. and {Edvardsson}, B. and {Eriksson}, K., et al., 2008, A\&A, 486, 951G
\bibitem[Hill et al.(2022a)]{Hillmainparam}Hill, L., Thomas, D.,  Maraston C., et al., 2022, MNRAS, 509, 4308H (Hill et al. 2022a)
\bibitem[Hill et al.(2022b)]{Hillalphaparam}Hill, L., Thomas, D. Maraston C., et al., 2022, MNRAS, 517, 4275H (Hill et al. 2022b)
\bibitem[Holtzman et al.(2018)]{Holtzman18}{Holtzman}, Jon A.,  {Hasselquist}, Sten, {Shetrone}, Matthew, et al., 2018, AJ, 156, 125H
\bibitem[Imig et al.(2022)]{Imigparam}Imig, J., Holtzman, H., Yan, R., et al. 2022, AJ, 163, 56
\bibitem[Ivanov et al.(2019)]{Ivanov19} vanov, Valentin D.; Coccato, Lodovico; Neeser, Mark J. et al., 2019, A\&A, 629A, 100I
\bibitem[J{\"o}nsson et al.(2018)]{Jonsson18}{J{\"o}nsson}, H., {Allende Prieto}, C., {Holtzman}, J. A., et al. 2018, AJ, 156, 126J
\bibitem[J{\"o}nsson et al.(2020)]{Jonsson20}{J{\"o}nsson}, H., {Holtzman}, J. A., {Allende Prieto}, C., et al., 2020, AJ, 160, 120J 
\bibitem[Koleva et al.(2008)]{Koleva08}{Koleva}, M., {Prugniel}, P.,  {Ocvirk}, P., {Le Borgne}, D., {Soubiran}, C., 2008, MNRAS, 385, 1998K
\bibitem[Kroupa(2001)]{Kroupa01}Kroupa P., 2001, MNRAS, 322, 231
\bibitem[Kroupa(2002)]{Kroupa02}Kroupa P., 2002, Science, 295, 82
\bibitem[Kroupa et al.(2013)]{Kroupa13} Kroupa P., Weidner C., Pflamm-Altenburg J., Thies I., Dabringhausen J.,
Marks M., Maschberger T., 2013, Planets, Stars and Stellar Systems.
Volume 5: Galactic Structure and Stellar Populations, p. 115
\bibitem[Kurucz(1979)]{kurucz79} Kurucz, R. L. 1979, \apjs, 40, 1
\bibitem[Kurucz \& Avrett(1981)]{kurucz81} Kurucz, R. L., \& Avrett, E. H. 1981, SAO Special Report, 391
\bibitem[Lan{\c{c}}on \& Wood(2000)]{Lancon2000}Lan{\c{c}}on, A., Wood, P. R., 2000, A\&AS, 146, 217
\bibitem[Law et al.(2016)]{Law16}Law, D. R., Cherinka, B., Yan, R., et al. 2016, AJ, 152, 83
\bibitem[Law et al.(2021)]{Law21} {Law}, David R., {Westfall}, Kyle B., {Bershady}, Matthew A. et al., 2021, AJ, 161, 52L
\bibitem[Lazarz et al.(2022)]{Lazarz22}{Lazarz}, Daniel,  {Yan}, Renbin,  {Wilhelm}, Ronald et al., 2022, A\&A, 668A, 21L
\bibitem[Le Borgne et al.(2003)]{stelibref}{Le Borgne}, J.-F., {Bruzual}, G., {Pell{\'o}}, R., et al. 2003, A\&A, 402, 433L
\bibitem[Leitherer et al.(2010)]{Leitherer10}{Leitherer}, Claus, {Ortiz Ot{\'a}lvaro}, Paula A. et al., 2010, ApJS, 189, 309L
\bibitem[Mainieri et al.(2024)]{Mainieri24}Mainieri, Vincenzo,   Anderson, Richard I.,  Brinchmann, Jarle, et al., 2024, arXiv, 240305398 
\bibitem[Majewski et al.(2017)]{Majewski17}{Majewski}, Steven R.,  {Schiavon}, Ricardo P., {Frinchaboy}, Peter M., 2017, AJ, 154, 94M 
\bibitem[Maraston(2005)]{Maraston05} Maraston, Claudia, 2005, MNRAS, 362, 799M
\bibitem[Marigo et al.(2017)]{Marigo17}{Marigo}, Paola, {Girardi}, L{\'e}o,  {Bressan}, Alessandro et al., 2017, ApJ, 835, 77M
\bibitem[Martins et al.(2005)]{Martins05} Martins, L. P.,  {Gonz{\'a}lez Delgado}, R. M., Leitherer, C., {Cervi{\~n}o}, M., \& Hauschildt, P.,  2005, MNRAS, 358, 49M
\bibitem[M{\'e}sz{\'a}ros et al.(2012)]{meszaros12} M{\'e}sz{\'a}ros, S., Allende Prieto, C., Edvardsson, B., et al. 2012, AJ, 144, 120
\bibitem[Munari et al.(2005)]{Munari05}Munari, U., Sordo, R., Castelli, F., Zwitter, T. 2005, A\&A, 442,1127M
\bibitem[Partridge \& Schwenke(1997)]{partridge97} Partridge, H., \& Schwenke, D. W. 1997, J. Chem. Phys., 106, 4618
\bibitem[Pickles(1985)]{Pickles85}{Pickles}, A.~J., 1985, ApJS, 59, 33P
\bibitem[Pickles(1998)]{Pickles98}{Pickles}, A.~J., 1998, PASP, 110, 863P
\bibitem[Prugniel \& Soubiran(2001)]{elodie}Prugniel, Ph., Soubiran, C. 2001, A\&A, 369, 1048P
\bibitem[Randich et al.(2022)]{Randich22}{Randich}, S., {Gilmore}, G., {Magrini}, L. et al., 2022A\&A, 666A, 121R
\bibitem[Rayner et al.(2009)]{Rayner09}Rayner, J. T., Cushing, M. C., Vacca, W. D. 2009, ApJS, 185, 289
\bibitem[S{\'a}nchez-Bl{\'a}zquez et al.(2006)]{milesref}{S{\'a}nchez-Bl{\'a}zquez}, P., {Peletier}, R.~F.,     {Jim{\'e}nez-Vicente}, J., et al. 2006, MNRAS, 371, 703S	
\bibitem[Sbordone et al.(2004)]{sbordone04} Sbordone, L., Bonifacio, P., Castelli, F., \& Kurucz, R. L. 2004, Memorie della Societa
Astronomica Italiana Supplementi, 5, 93
\bibitem[Schwenke(1998)]{schwenke98} Schwenke, D. W. 1998, Faraday Discussions, 109, 321
\bibitem[Schlegel et al.(1998)]{Schlegel98}{Schlegel}, D.~J., {Finkbeiner}, D.~P., {Davis}, M.
\bibitem[Silva \&Cornell(1992)]{Silva92}Silva, D. R., Cornell, M. E., 1992, ApJS, 81, 865, doi: 10.1086/191706
\bibitem[Smee et al.(2013)]{Smee13}{Smee}, Stephen A., {Gunn}, James E., {Uomoto}, Alan, et al., 2013, AJ, 146, 32
\bibitem[Tang et al.(2014)]{Tang14}{Tang}, Jing,  {Bressan}, Alessandro,  {Rosenfield}, Philip et al., 2014, MNRAS, .445, 4287T
\bibitem[Valdes et al.(2004)]{Valdes04}Valdes, F., Gupta, R., Rose, J. A., Singh, H. P., Bell, D. J., 2004, ApJS, 152, 251, doi: 10.1086/386343
\bibitem[Verro et al.(2022)]{dr3xsl}Verro, K., Trager, S. C., Peletier, R. F., et al. 2022, A\&A, 660, A34
\bibitem[Villaume et al.(2017)]{Villaume17}Villaume, A., Conroy, C., Johnson, B., et al. 2017, ApJS, 230, 23
\bibitem[Worthey et al.(1994)]{Worthey94}Worthey, G., Faber, S. M., Gonzalez, J. J., Burstein, D., 1994, ApJS, 94, 687
\bibitem[Yan et al.(2016)]{Yan16}Yan, R., Tremonti, C., Bershady, M. A., et al. 2016, AJ, 151, 8Y
\bibitem[Yan et al.(2019)]{Yan19}Yan, R., Chen, Y.-P., Lazarz, D., et al., 2019, ApJ, 883, 175Y 
\bibitem[Yan et al.(2026 in prep)]{Yan24}Yan, et al., 2026, in prep.
\bibitem[Zhao et al.(2012)]{zhao12}Zhao, G., Zhao, Y.-H., Chu, Y.-Q., Jing, Y.-P., Deng, L.-C. 2012, Research in Astronomy and Astrophysics, 12, 723, doi: 10.1088/1674-4527/12/7/002
\bibitem[Zwitter et al.(2004)]{Zwitter04}Zwitter, T., Castelli, F., Munari, U. 2004, A\&A, 417, 1055

\end{thebibliography}
\end{document}